\documentclass[10pt,english,a4paper]{article}

\makeatletter
\renewcommand*{\@fnsymbol}[1]{\ensuremath{\ifcase#1\or 1\or 2\or 3\or
   \mathsection\or \mathparagraph\or \|\or **\or \dagger\dagger
   \or \ddagger\ddagger \else\@ctrerr\fi}}
\makeatother

\usepackage{ifpdf}   
\usepackage{multido} 
\usepackage{ifthen}

\usepackage{color}

\ifpdf
\usepackage[pdftex]{graphicx}
\else
\usepackage[dvips]{graphicx}
\fi

\usepackage{float,floatflt,afterpage}
\usepackage{wrapfig}
\usepackage{placeins}

\usepackage{etex}
\usepackage{tikz}
\usetikzlibrary{snakes}
\usepackage{textcomp}

\usepackage[arrow,matrix,curve]{xy} 
\usepackage{subfigure} 

\graphicspath{{./figures/}}

\usepackage{babel}
\usepackage[latin1]{inputenc}

\usepackage[centertags,intlimits]{amsmath}
\usepackage{amsthm}
\usepackage{amssymb}
\usepackage{mathtools}
\usepackage{mathrsfs}
\usepackage[normalem]{ulem}

\theoremstyle{break}
\allowdisplaybreaks[1]

\usepackage{exscale}
\usepackage{dsfont}
\usepackage{bbm}

\usepackage[numbers,square,sort&compress]{natbib}

\ifpdf
\usepackage[pdftex,
  bookmarks,
  bookmarksopen=true,
  bookmarksnumbered=true,
  backref,
  linkcolor=blue,
  urlcolor=blue,
  pdfauthor={Andreas Fischle},
  pdftitle={The relaxed-polar mechanism of locally optimal Cosserat rotations for an idealized nanoindentation and comparison with 3D-EBSD experiments}
  hyperfigures=true,
  pdfpagelabels,
  pagebackref,
  hypertexnames=true,
  plainpages=false,
  naturalnames]{hyperref}%
\else
\usepackage[dvips,breaklinks]{hyperref} 
\usepackage{breakurl}
\fi

\usepackage{hypernat}

\usepackage{url} 
\usepackage{breakurl} 

\makeatletter
\def\url@leostyle{%
  \@ifundefined{selectfont}{\def\UrlFont{\sf}}{\def\UrlFont{\small\ttfamily}}}
\makeatother
\usepackage{enumerate}
\usepackage{verbatim}        
\usepackage{lscape}          
\usepackage{array,longtable} 
\usepackage{listings}        
\usepackage{caption}

\usepackage{makeidx}

\newcommand{\Lc}{L_{\rm c}}
\def\and{{\rm and}}          
\def\det#1{{\rm det}\,{#1}}  

\def\exp{{\rm exp}}          

\def\skew{{\rm skew}}        

\newfont{\Sf}{cmssbx10 scaled 2074}
\newbox{\assem}
\savebox{\assem}(15,2)[bl]{
\begin{normalsize}
\unitlength=1.0mm
\begin{picture}(6.00,0.00)
\put(.0,1.0){\makebox(0,0)[cc]{$\cup$}}
\put(.0,-1.0){\makebox(0,0)[cc]{$\mbox{\scriptsize \it n=1}$}}
\put(.0,4.0){\makebox(0,0)[cc] {\scriptsize $\it n_{\mbox{\tiny\it elm}}$}}
\end{picture}
\end{normalsize}}

\newbox{\asse}
\savebox{\asse}(15,2)[bl]{
\begin{normalsize}
\unitlength=1.0mm
\begin{picture}(8.00,0.00)
\put(.0,2.0){\makebox(0,0)[cc]{\Sf A}}
\end{picture}
\end{normalsize}}

\def\sqtwo3{{\textstyle {\sqrt{2 \over 3}}}}   
\newcommand{\IP}{{\rm I\kern-.18em P}}           
\newcommand{\II}{{\rm I\kern-.18em I}}           
\newcommand{\IF}{{\rm I\kern-.25em F}}           
\newcommand{\IE}{{\rm I\kern-.25em E}}           
\def\IR{{\rm I\kern-.15em R}}

\newcommand{\ia}{{\rm\kern.24em                  
   \vrule width.02em height0.9ex depth-.05ex
   \kern-.26em a}}
\newcommand{\ic}{{\rm\kern.24em                  
   \vrule width.02em height0.9ex depth-.05ex
   \kern-.26em c}}
\newcommand{\IA}{{\rm\kern.22em                  
    \vrule width.02em
        height0.5ex depth 0ex
    \kern-.24em A}}
\newcommand{\IC}{{\rm\kern.24em                  
   \vrule width.02em height1.4ex depth-.05ex
   \kern-.26em C}}
\DeclareMathOperator{\sign}{sign}
\DeclareMathOperator{\dist}{dist}

\newcommand{\Dphi}{\nabla\! \mathbf{\varphi}}

\renewcommand{\epsilon}{\varepsilon}

\newcommand{\dV}{\,{\rm dV} }

\DeclareMathOperator{\Curl}{Curl}

\newcommand{\norm}[1]{\|#1\|}
\newcommand{\abs}[1]{\left| #1 \right|}

\newtheorem{lem}{Lemma}[section]

\newtheorem{rem}[lem]{Remark}
\newtheorem{defi}[lem]{Definition}

\newtheorem{theo}[lem]{Theorem}

\newtheorem{cor}[lem]{Corollary}
\newtheorem{exa}[lem]{Example}
\newtheorem{prob}[lem]{Problem}
\newtheorem{propo}[lem]{Proposition}

\newcommand{\deref}[1]{Definition \ref{#1}}

\newcommand{\probref}[1]{Problem \ref{#1}}
\newcommand{\proporef}[1]{Proposition \ref{#1}}

\newcommand{\R}{\mathbb{R}}
\newcommand{\N}{\mathbb{N}}

\renewcommand{\S}{\mathbb{S}}
\DeclareMathOperator{\SL}{SL}
\DeclareMathOperator{\GL}{GL}
\DeclareMathOperator{\SO}{SO}
\let\O\undefined
\DeclareMathOperator{\O}{O}

\DeclareMathOperator{\skewop}{skew}
\renewcommand{\skew}{\skewop}
\DeclareMathOperator{\diag}{diag}
\DeclareMathOperator{\sym}{sym}
\DeclareMathOperator{\Tr}{tr}

\DeclareMathOperator{\dev}{dev}

\DeclareMathOperator{\polar}{polar}

\newcommand{\Sym}{ {\rm{Sym}} }
\newcommand{\Psym}{ {\rm{Sym^+}} }
\newcommand{\id}{{\boldsymbol{\mathbbm{1}}}}

\DeclareMathOperator{\Det}{det}

\renewcommand{\det}[1]{ {\Det[{#1}]} }
\newcommand{\tr}[1]{ {\Tr \left[{#1}\right]} }

\newcommand{\elastic}[1]{{#1}_{\mathrm{e}}}
\newcommand{\plastic}[1]{{#1}_{\mathrm{p}}}

\newcommand{\micro}[1]{{#1}_{\mathrm{micro}}}

\newcommand{\Fe}{\elastic{F}}
\renewcommand{\Re}{\elastic{R}}
\newcommand{\Ue}{\elastic{U}}
\newcommand{\Ve}{\elastic{V}}

\newcommand{\Fp}{\plastic{F}}
\newcommand{\Rp}{\plastic{R}}
\newcommand{\Up}{\plastic{U}}

\newcommand{\mumic}{\micro{\mu}}
\newcommand{\kappamic}{\micro{\kappa}}

\newcommand{\mue}{\elastic{\mu}}
\newcommand{\kappae}{\elastic{\kappa}}

\newcommand{\secref}[1]{Section \ref{#1}}
\newcommand{\figref}[1]{Figure \ref{#1}}

\definecolor{orange}{rgb}{1.0,0.5,0}

\DeclareMathOperator{\Reals}{\mathbb{R}}
\renewcommand{\R}{\Reals}

\DeclareMathOperator{\argminmathop}{arg\,min}
\DeclareMathOperator{\arginfmathop}{arg\,inf}

\newcommand{\argmin}[2]{\mathchoice{\underset{#1}{\argminmathop}\, {#2}}{\argminmathop_{#1}\, {#2}}{}{}}
\newcommand{\arginf}[2]{\mathchoice{\underset{#1}{\arginfmathop}\, {#2}}{\arginfmathop_{#1}\, {#2}}{}{}}

\newcommand{\mathematica}{\texttt{Mathematica}}

\newcommand{\restrict}[2]{\left.#1\right\rvert_{#2}\,}
\newcommand{\scalprod}[2]{\big<#1,\,#2\big>}

\newcommand{\setdef}[2]{\lbrace #1 \;\vert\; #2\rbrace}

\DeclareMathOperator{\spanop}{span}

\newcommand{\vspan}[1]{\spanop\left(\left\{ #1 \right\}\right)}

\newcommand{\closure}[1]{\overline{#1}}

\newcommand{\hsnorm}[1]{\left\lVert #1 \right\rVert}

\DeclareMathOperator{\RPosZ}{\sideset{}{_0^+}\Reals}

\newcommand{\eqdef}{\coloneqq}
\newcommand{\defeq}{\eqqcolon}
\newcommand{\eqiso}{\cong}
\newcommand{\isequivto}{\Longleftrightarrow}

\newcommand{\pu}[1]{\,\text{#1}}

\DeclarePairedDelimiter\floor{\lfloor}{\rfloor} 

\newcommand{\mstretch}{\overline{U}}

\DeclareMathOperator{\rpolar}{rpolar}

\newcommand{\ump}{u^{\rm mmp}}
\newcommand{\smp}{s^{\rm mmp}}

\DeclareMathOperator{\sradmm}{\rho_{\mu,\,\mu_c}}

\DeclareMathOperator{\wmm}{W_{\mu,\mu_c}}
\DeclareMathOperator{\wsym}{W_{1,0}}

\newcommand{\domc}{\mathrm{D}^\mathrm{C}}
\newcommand{\domn}{\mathrm{D}^\mathrm{NC}}

\newcommand{\milDir}[1]{{\left[{#1}\right]}}

\newcommand{\milPlane}[1]{{\left({#1}\right)}}
\newcommand{\milPlaneFamily}[1]{{\left\{{#1}\right\}}}
\newcommand{\milSlipSystem}[2]{{\left({#1}\right)\!\left[{#2}\right]}}

\newcommand{\angstrom}{\AA}

\newcommand{\removedAccepted}[1]{}

\newcommand{\countres}{
  \setcounter{equation}{0}
  \setcounter{figure}{0}
  \setcounter{table}{0}
}

\renewcommand{\baselinestretch}{1.0}          
\makeatletter

\renewcommand{\itemize}{%
  \ifnum \@itemdepth >\thr@@\@toodeep\else
    \advance\@itemdepth\@ne
    \edef\@itemitem{labelitem\romannumeral\the\@itemdepth}%
    \expandafter
    \list
      \csname\@itemitem\endcsname
      {\def\makelabel##1{\hss\llap{##1}}%
        \topsep=.8ex\itemsep=-.2ex}%
  \fi}

\renewcommand\section{\@startsection {section}{1}{\z@}%
  {-3.5ex \@plus -1ex \@minus -.2ex}%
  {2.3ex \@plus.2ex}%
  {\boldmath\normalfont\Large\bfseries}}
\renewcommand\subsection{\@startsection{subsection}{2}{\z@}%
  {-3.25ex\@plus -1ex \@minus -.2ex}%
  {1.5ex \@plus .2ex}%
  {\boldmath\normalfont\large\bfseries}}
\renewcommand\subsubsection{\@startsection{subsubsection}{3}{\z@}%
  {-3.25ex\@plus -1ex \@minus -.2ex}%
  {1.5ex \@plus .2ex}%
  {\boldmath\normalfont\normalsize\bfseries}}
\renewcommand\paragraph{\@startsection{paragraph}{4}{\z@}%
  {3.25ex \@plus1ex \@minus.2ex}%
  {-1em}%
  {\boldmath\normalfont\normalsize\bfseries}}
\renewcommand\subparagraph{\@startsection{subparagraph}{5}{\parindent}%
  {3.25ex \@plus1ex \@minus .2ex}%
  {-1em}%
  {\boldmath\normalfont\normalsize\bfseries}}

\makeatother
 
\title{The relaxed-polar mechanism of locally optimal
  Cosserat rotations for an idealized nanoindentation
  and comparison with 3D-EBSD experiments}

\author{Andreas Fischle
\!\!\footnote{Corresponding author: Andreas Fischle,
  Institut f\"ur Numerische Mathematik
  TU Dresden,
  Zellescher Weg 12-14,
  01069 Dresden,
  Germany,
  email: andreas.fischle@tu-dresden.de\newline
  Dresden Center for Computational Materials Science (DCMS) (http://dcms.tu-dresden.de/)
}
\;,\quad
Patrizio Neff
\!\footnote{Patrizio Neff,
Head of Lehrstuhl f\"{u}r Nichtlineare Analysis und Modellierung,
Fakult\"{a}t f\"{u}r Mathematik,
Universit\"{a}t Duisburg-Essen,
Thea-Leymann Str. 9,
45127 Essen,
Germany,
email: patrizio.neff@uni-due.de}
\;,\quad and \quad
Dierk Raabe
\!\!\footnote{Dierk Raabe,
\removedAccepted{CEO} Max-Planck-Institut f\"ur Eisenforschung,
Max-Planck-Str. 1,
40237 D\"usseldorf,
Germany,
email: d.raabe@mpie.de
}
}

\begin{document}

\selectfont
\maketitle

\pagenumbering{arabic}

\begin{abstract}
    The rotation $\polar(F) \in \SO(3)$ arises as the unique orthogonal
    factor of the right polar decomposition $F = \polar(F)\,U$ of a
    given invertible matrix $F \in \GL^+(3)$. In the context of
    nonlinear elasticity Grioli (1940) discovered a geometric variational
    characterization of $\polar(F)$ as a unique energy-minimizing
    rotation. In preceding works, we have analyzed a generalization of
    Grioli's variational approach with weights (material parameters)
    $\mu > 0$ and $\mu_c \geq 0$ (Grioli: $\mu = \mu_c$). The energy
    subject to minimization coincides with the Cosserat shear--stretch
    contribution arising in any geometrically nonlinear, isotropic and
    quadratic Cosserat continuum model formulated in the deformation
    gradient field $F \eqdef \nabla\varphi: \Omega \to \GL^+(3)$ and
    the microrotation field $R: \Omega \to \SO(3)$. The corresponding
    set of non-classical energy-minimizing rotations
    \begin{align*}
      \rpolar^\pm_{\mu,\mu_c}(F) \eqdef \argmin{R\,\in\,\SO(3)}{\Big\{\wmm(R\,;F)
        \eqdef
        \mu  \hsnorm{\sym(R^TF - \id)}^2 + \mu_c\hsnorm{\skew(R^TF - \id)}^2\Big\}}
    \end{align*}
    represents a new \emph{relaxed-polar} mechanism. Our goal is to motivate
    this mechanism by presenting it in a relevant setting. To this end, we
    explicitly construct a deformation mapping $\varphi_{\rm nano}$
      which models an idealized nanoindentation
    and compare the corresponding optimal rotation patterns
    $\rpolar^\pm_{1,0}(F_{\rm nano})$ with experimentally obtained 3D-EBSD
    measurements of the disorientation angle of lattice rotations due to
    a nanoindentation in solid copper. We observe that the
    non-classical \emph{relaxed-polar} mechanism can produce interesting
    counter-rotations. A possible link between Cosserat theory and finite
    multiplicative plasticity theory on small scales is also explored.
\end{abstract}

\vspace*{1.125cm}
{\small

\vspace*{0.125cm}
\textbf{AMS 2010:}
  15A24,   
  22E30,   
  74A30,   
  74A35,   
  74B20,   
  74E15,   
  74G65,   
  74N05,   
  74N15,   
  82D25.   
} 
 \countres

\newpage

\setcounter{tocdepth}{3}
\renewcommand{\baselinestretch}{-1.0}\normalsize
{
  \small
  \tableofcontents
}
\renewcommand{\baselinestretch}{1.0}\normalsize

\section{Introduction}\label{sec:intro}
We consider the weighted optimality problem for the Cosserat
shear--stretch energy $\wmm:\; \SO(3) \,\times\, \GL^+(3) \to \RPosZ$,\\
\begin{equation}
  \wmm(R\,;F) \;\eqdef\; \mu\, \hsnorm{\sym(R^TF - \id)}^2
  \,+\,
  \mu_c\,\hsnorm{\skew(R^TF - \id)}^2\;.
  \label{eq:intro:wmm}
\end{equation}
The arguments are the deformation gradient field
$F \eqdef \nabla\varphi: \Omega \to \GL^+(3)$ induced by a deformation
mapping $\varphi: \Omega \to \Omega_{\rm def} \eqdef \varphi(\Omega)$
and the microrotation field $R: \Omega \to \SO(3)$ evaluated at a given
point of the domain $\Omega$. Further, we use the notation
$\sym(X) \eqdef \frac{1}{2}(X + X^T)$,
$\skew(X) \eqdef \frac{1}{2}(X - X^T)$,
$\scalprod{X}{Y} \eqdef \tr{X^TY}$
and we denote the induced Frobenius matrix norm by
$\hsnorm{X}^2 \eqdef \scalprod{X}{X} = \sum_{1 \leq i,j \leq n} X_{ij}^2$.
The weight $\mu > 0$ coincides with the Lam\'e shear modulus from linear
elasticity and the weight $\mu_c \geq 0$ can be identified with the
so-called Cosserat couple modulus.

The energetic contribution~\eqref{eq:intro:wmm} arises in any geometrically
nonlinear, isotropic and quadratic Cosserat micropolar continuum model. Note
that the local energy contribution in a Cosserat model can \emph{always} be
expressed as $W = W(\mstretch)$, i.e., as a function of the first Cosserat
deformation tensor $\mstretch \eqdef R^TF$. This structure is implied by
objectivity requirements and was for the first time observed by the Cosserat
brothers~\cite[p.~123, eq.~(43)]{Cosserat09}, see also~\cite{Eringen99}
and~\cite{Maugin:1998:STPE}. Since $\mstretch$ is a non-symmetric quantity,
the most general isotropic and quadratic expression for the local energy
contribution which vanishes identically at the reference state is of the
form\footnote{Note that the Cosserat brothers never proposed any specific
  expression for the local energy $W = W(\mstretch)$, i.e.,
  they never proposed a constitutive material model inside their original
  Cosserat framework, as Eringen correctly observes in the introduction
  of~\cite{Eringen99}. Linear Cosserat theory was developed only some 50
  years later. The particular quadratic ansatz for $W = W(\mstretch)$
  considered here, is motivated by a direct extension of the quadratic
  energy in the linear theory of Cosserat models, see, e.g.~\cite{Jeong:2009:NLIC,Neff_Jeong_bounded_stiffness09,Neff_Jeong_Conformal_ZAMM08}. A
  physical motivation for~\eqref{eq:intro:wmm} with $\mu_c = 0$ will be
  given in~\secref{sec:plasticity}.}
\begin{equation}
  \label{intro:wgeneral}
  \underbrace{\mu\, \hsnorm{\sym(\mstretch - \id)}^2
    \,+\,
    \mu_c\,\hsnorm{\skew(\mstretch - \id)}^2}_{\text{``shear--stretch energy''}}
  \quad+\quad
  \underbrace{\frac{\lambda}{2}\,\tr{\mstretch - \id}^2}_{\text{``linearized volumetric energy''}}\;.
\end{equation}
The parameter $\lambda$ can be identified with the second Lam\'e
parameter. In the following, we dispense with the corresponding term,
since it couples the rotational and volumetric response; a feature not
present in the isotropic, geometrically linear Cosserat model.

In~\cite{Fischle:2016:OC2D}, we have proved a still surprising reduction
lemma~\cite[Lem.~2.2, p.~4]{Fischle:2016:OC2D} for the material parameters
(weights) $\mu$ and $\mu_c$ which is valid
for \emph{all} space dimensions $n \geq 2$. This lemma singles out a
\emph{classical parameter range} $\mu_c \geq \mu > 0$ and a
\emph{non-classical parameter range} $\mu > \mu_c \geq 0$ for $\mu$
and $\mu_c$ and reduces both ranges to an associated limit case. The
\emph{classical limit case} is given by $(\mu,\mu_c) = (1,1)$ and the
\emph{non-classical limit case} is given by $(\mu,\mu_c) = (1,0)$.
Exploiting the parameter
reduction~\cite[Lem.~2.2, p.~4]{Fischle:2016:OC2D}
for~\probref{intro:prob_wmm} below, we were able to discuss and
classify the solutions in dimension $n = 2$. Thus a crucial observation
was made: the classical and the non-classical parameter ranges for
$\mu$ and $\mu_c$ characterize a classical and a non-classical regime
for the optimal Cosserat rotations. Most importantly, the latter
allows for interesting new rotation patterns.

Subsequently, in~\cite{Fischle:2016:OC3D}, we have applied computer
algebra to the challenging three-dimensional
\begin{prob}[Weighted optimality in dimension $n = 3$]
  Let $\mu > 0$ and $\mu_c \geq 0$. Compute the set of optimal rotations
  \label{intro:prob_wmm}
  \begin{equation}
    \argmin{R\,\in\,\SO(3)}{\wmm(R\,;F)}
    \;\eqdef\;
    \argmin{R\,\in\,\SO(3)}{
      \left\{
      \mu\, \hsnorm{\sym(R^TF - \id)}^2
      \,+\,
      \mu_c\,\hsnorm{\skew(R^TF - \id)}^2\right\}
    }
    \label{eq:intro:weighted_wmm}
  \end{equation}
  for given parameter $F \in \GL^+(3)$ with distinct singular values
  $\sigma_1 > \sigma_2 > \sigma_3 > 0$.
\end{prob}
The resulting explicit expressions allowed us to extract a non-classical
geometric mechanism represented by the non-classical minimizers. It is
the principal goal of the current paper to study
this \emph{relaxed-polar} mechanism in an interesting scenario.

In what follows $\polar(F) \in \SO(3)$ denotes the unique orthogonal
factor of the right polar decomposition $F = \polar(F)\,U(F)$, where
$F \eqdef \nabla\varphi \in \GL^+(3)$ is the deformation gradient.
The right Biot-stretch tensor $U(F) \eqdef \sqrt{F^TF} \in \Psym(3)$
is positive definite symmetric. Furthermore, we make the assumption
that the singular values of $F \in \GL^+(3)$ satisfy
$\sigma_1 > \sigma_2 > \sigma_3 > 0$. Recall that the singular
values of $F \in \GL^+(3)$ are defined as the eigenvalues of
$U = \sqrt{F^TF} \in \Psym(3)$.

As suggested by the nomenclature, the polar factor $\polar(F)$ is
the \emph{unique} minimizer for~\eqref{eq:intro:wmm} in the
\emph{classical} parameter range $\mu_c \geq \mu > 0$, for all
$n \geq 2$. For this generalized version of Grioli's theorem,
see~\cite{Grioli40,Guidugli:1980:EPP,Neff_Grioli14},
or~\cite[Cor.~2.4, p.~5]{Fischle:2016:OC2D}. This variational
characterization of the polar factor inspired us to introduce the
following
\begin{defi}[Relaxed polar factor(s)]
  Let $\mu > 0$ and $\mu_c \geq 0$. We denote the set-valued mapping that assigns
  to a given parameter $F \in \GL^+(3)$ its associated set of
  energy-minimizing rotations by
  \begin{equation*}
    \rpolar_{\mu,\mu_c}(F) \quad \eqdef\quad \argmin{R\,\in\,\SO(3)}{\wmm(R\,;F)}\;.
  \end{equation*}
\end{defi}

By now, the classical parameter domain $\mu_c \geq \mu > 0$ is very well
understood. This allows us to focus entirely on the non-classical parameter
range $\mu > \mu_c \geq 0$ in our efforts to solve~\probref{intro:prob_wmm}.
Furthermore, the parameter reduction
lemma~\cite[Lem.~2.2, p.~4]{Fischle:2016:OC2D}
states that it is sufficient to solve the non-classical limit case
$(\mu,\mu_c) = (1,0)$, since it implies the following equivalence
for all $n \geq 2$:
\begin{equation}
  \argmin{R\,\in\,\SO(3)}{W_{\mu,\mu_c}(R\,;F)}
  \quad=\quad
  \argmin{R\,\in\,\SO(3)}{W_{1,0}(R\,; \widetilde{F}_{\mu,\mu_c})}\;.
\end{equation}
On the right hand side, we notice a \emph{rescaled deformation gradient}
$$\widetilde{F}_{\mu,\mu_c} \eqdef \lambda^{-1}_{\mu,\mu_c} \, F \in \GL^+(3)$$
which is obtained from $F \in \GL^+(3)$ by multiplication with the inverse of
the \emph{induced scaling parameter}
$\lambda_{\mu,\mu_c} \eqdef \frac{\mu}{\mu - \mu_c} > 0$. We note that we use
the previous notation throughout the text and further introduce
the \emph{singular radius} $\rho_{\mu,\mu_c} \eqdef \frac{2\mu}{\mu - \mu_c}$.

It follows that the set of optimal Cosserat rotations can be described by
\begin{equation}
  \rpolar_{\mu,\mu_c}(F) \;=\; \rpolar_{1,0}(\widetilde{F}_{\mu,\mu_c})
\end{equation}
for the entire non-classical parameter range $\mu > \mu_c \geq 0$.
Based on our derivation presented in~\secref{sec:plasticity},
it is for the most part sufficient to focus our discussion on the
distinguished non-classical limit case $\mu_c = 0$. As it turns out,
there are at most two non-classical minimizing branches
of rotations $\rpolar_{\mu,\mu_c}^\pm(F)$ in the non-classical
parameter range which we can distinguish by a sign; see \eqref{eq:rpolar10pm} for the definition. Note that for every shear modulus $\mu > 0$
and fixed choice of sign, we have
\begin{equation}
  \rpolar^\pm_{\mu,0}(F)
  \;=\;
  \rpolar^\pm_{1,0}(F).
\end{equation}

\subsection{The locally energy-minimizing Cosserat rotations $\rpolar^\pm_{\mu,\mu_c}(F)$}
We briefly present the geometric characterization of the optimal
Cosserat rotations $\rpolar^\pm_{\mu,\mu_c}(F)$ obtained
in~\cite{Fischle:2016:OC3D}. Let $R \in \SO(n)$ for $n = 2,3$ and
let $\S^2 \subset \R^3$ denote the unit $2$-sphere. In dimension
$n = 3$, we use the well-known angle-axis parametrization of
rotations which we write as $[\alpha,\, r]$ and which allows to
parametrize $R \in \SO(3)$.\footnote{The angle-axis parametrization
  is singular, but this is not an issue for our exposition.}
Here the rotation angle is $\alpha \in (-\pi,\pi]$ and $r \in \S^{2}$
specifies the oriented rotation axis.

In order to reduce the parameter space $\GL^+(3)$, we use the (unique)
polar decomposition\footnote{For an introduction to the polar
  and singular value decomposition, see, e.g.,~\cite{DSerre02} and for
  recent related results on fundamental variational characterizations
  of the polar factor $\polar(F)$, see~\cite{Neff_Grioli14,Lankeit:2014:MML,Neff:2014:LMP}
  and references therein.}
\mbox{$F = \polar(F)\,U$} and the (non-unique) spectral
decomposition of $U =\sqrt{F^TF} \in \Psym(3)$ given by
$U = QDQ^T$, $Q \in \SO(3)$, and expand
\begin{equation}
  R^TF = R^T\polar(F) U = R^T\polar(F) QDQ^T\;.
\end{equation}
Here, the diagonal matrix $D = \diag(\sigma_1,\sigma_2,\sigma_3)$
contains the eigenvalues of $U$ on its diagonal. These are, by
definition, the singular values of $F \in \GL^+(3)$. Note that this
is a particular form of the singular value decomposition (SVD).
If, furthermore, $F$ has only simple singular values, then it is
always possible to choose the rotation $Q \in \SO(3)$ such that an
ordering $\sigma_1 > \sigma_2 > \sigma_3 > 0$ is achieved.

Exploiting that $Q \in \SO(3)$, we now transform the Cosserat
shear--stretch energy into principal axis coordinates. This
simplification makes use of the isotropy of the energy. For
the actual computation, note first that
\begin{align}
  &Q^T(\sym(R^TF) - \id)Q = Q^T\left(\sym(R^T\polar(F) QDQ^T) - \id)\right)Q  \label{eq:symtransform}\\
  &\quad=\sym(Q^TR^T\polar(F) QDQ^TQ - Q^TQ) = \sym(\underbrace{Q^TR^T\polar(F) Q}_{\defeq\;\widehat{R}}D - \id) = \sym(\widehat{R}D - \id)\;.\notag
\end{align}
In the process, it is natural to introduce a \emph{relative} rotation
\begin{equation}
  \label{eq:Rhat}
  \widehat{R} \eqdef Q^TR^T\polar(F) Q
\end{equation}
which acts relative to the continuum rotation $\polar(F)$. The action
of $\widehat{R}$ is defined with respect to the coordinate system
induced by the columns of $Q$, i.e., in a positively oriented frame of
principal directions of $U$. This interpretation is also underligned by
the inverse formula
\begin{equation}
  \label{eq:R}
  R = \left(Q\widehat{R}Q^T\polar(F) ^T\right)^T = \polar(F) \, Q\widehat{R}^TQ^T\;.
\end{equation}
The last relation allows us to recover the original absolute
rotation $R$ from the relative rotation $\widehat{R}$. Our next step
is to insert the transformed symmetric part~\eqref{eq:symtransform}
into the definition of
\begin{equation}
  W_{1,0}(R\,;F) =
  \hsnorm{\sym(R^TF - \id)}^2 =
  \hsnorm{Q^T\sym(R^TF - \id)Q}^2 =
  \hsnorm{\sym(\widehat{R}\,D - \id)}^2\label{eq:wsymtransformed}\;,
\end{equation}
where we have used that the conjugation by $Q^T$ preserves the
Frobenius matrix norm. Along the same lines, we can reduce the
shear-stretch energy for general $\mu > 0$ and $\mu_c \geq 0$,
as is easy to see. It follows, that it is sufficient to solve
for the relative rotation, i.e., we may consider
\begin{prob}[Diagonal form of weighted optimality in $n = 3$]
  \label{intro:prob_wmm_reduced}
  Let $\mu > 0$ and $\mu_c \geq 0$ and let
  $D = \diag(\sigma_1,\sigma_2,\sigma_3)$ with
  $\sigma_1 > \sigma_2 > \sigma_3 > 0$. Compute the set of
  optimal relative rotations
  \begin{equation}
    \argmin{\widehat{R}\,\in\,\SO(3)}{\wmm(\widehat{R}^T\,;D)}
    \;\eqdef\;
    \argmin{\widehat{R}\,\in\,\SO(3)}{
      \left\{
      \mu\, \hsnorm{\sym(\widehat{R}\,D - \id)}^2
      \,+\,
      \mu_c\,\hsnorm{\skew(\widehat{R}\,D - \id)}^2\right\}
    }
  \end{equation}
  for a given diagonal matrix $D$.
\end{prob}
We stress that the rotation angle of the
relative rotation $\widehat{R}$ is implicitly reversed
due to the correspondence $R^T \leftrightarrow \widehat{R}$.

The computation of the solutions to~\probref{intro:prob_wmm_reduced}
by computer algebra together with a statistical verification is one
of the core results obtained in~\cite{Fischle:2016:OC3D} which we
present next.

\begin{propo}[Energy-minimizing relative rotations for $(\mu,\mu_c) = (1,0)$]
  \label{propo:rhat}
  Let $\sigma_1 > \sigma_2 > \sigma_3 > 0$ be the singular values of
  $F \in \GL^+(3)$. Then the energy-minimizing relative rotations
  solving~\probref{intro:prob_wmm_reduced} are given by
  \begin{equation}
    \widehat{R}_{1,0}^{\pm}(F)
    \quad\eqdef\quad
    \begin{pmatrix}
      \cos \hat{\beta}^\pm_{1,0}  & -\sin \hat{\beta}^\pm_{1,0} & 0\\
      \sin \hat{\beta}^\pm_{1,0}  &  \cos \hat{\beta}^\pm_{1,0} & 0\\
      0                    &  0                   & 1\\
    \end{pmatrix}\;,
  \end{equation}
  where the optimal rotation angles $\hat{\beta}^\pm_{1,0} \in (-\pi,\pi]$
    are given by
    \begin{equation}
      \hat{\beta}^\pm_{1,0}(F)
      \quad\eqdef\quad
      \begin{cases}
        \; 0  &\quad,\text{if}\quad \sigma_1 + \sigma_2 \leq 2\;,\\
        \; \pm\arccos(\frac{2}{\sigma_1 + \sigma_2})
        &\quad,\text{if}\quad \sigma_1 + \sigma_2 \geq 2\;.
      \end{cases}
    \end{equation}
    Thus, in the non-classical regime $\sigma_1 + \sigma_2 \geq 2$,
    we obtain the explicit expression
    \begin{equation}
      \widehat{R}_{1,0}^{\pm}(F)
      \quad\eqdef\quad
      \begin{pmatrix}
        \frac{2}{\sigma_1 + \sigma_2}  & \mp \sqrt{1-\left(\frac{2}{\sigma_1 + \sigma_2}\right)^2} & 0\\
        \pm \sqrt{1-\left(\frac{2}{\sigma_1 + \sigma_2}\right)^2} & \frac{2}{\sigma_1 + \sigma_2} & 0 \\
        0 & 0 & 1
      \end{pmatrix}\;.
    \end{equation}
    In the classical regime $\sigma_1 + \sigma_2 \leq 2$, we simply
    obtain the relative rotation $\widehat{R}_{1,0}^{\pm}(F) = \id$,
    and there is no deviation from the polar factor $\polar(F)$
    at all.
\end{propo}
The corresponding \emph{absolute} representation of the optimal Cosserat
rotations is uniquely determined by the relation \eqref{eq:R}, which
leads us to define
\begin{equation}
  \label{eq:rpolar10pm}
  \rpolar^\pm_{1,0}(F) \; \eqdef \; \polar(F) \, Q(F)\left(\widehat{R}_{1,0}^\pm(F)\right)^TQ(F)^T\;.
\end{equation}%

Furthermore, due to the parameter reduction~\cite[Lem.~2.2]{Fischle:2016:OC2D},
it is always possible to recover the optimal rotations
$\rpolar^\pm_{\mu,\mu_c}(F)$ for general non-classical parameter choices
$\mu > \mu_c \geq 0$ from the non-classical limit case
$(\mu,\mu_c) = (1,0)$; cf.~\cite{Fischle:2016:OC2D}
and~\cite{Fischle:2016:OC3D} for details. For now, we shall defer
the explicit procedure, since it is quite instructive to interpret
the distinguished non-classical limit case $\mu = 1$ and
$\mu_c = 0$ first.

\subsection{Geometric-mechanical aspects of optimal Cosserat rotations}

It seems natural to introduce
\begin{defi}[Maximal mean planar stretch and strain]
  \label{defi:mmpss}
  Let $F \in \GL^+(3)$ with singular values
  $\sigma_1 \geq \sigma_2 \geq \sigma_3 > 0$. We introduce
  the \textbf{maximal mean planar stretch} $\ump$ and
  the \textbf{maximal mean planar strain} $\smp$ as follows:
  \begin{equation}
    \begin{aligned}
      \ump(F) &\;\eqdef\; \frac{\sigma_1 + \sigma_2}{2}\;,\quad\text{and}\\
      \smp(F) &\;\eqdef\; \frac{(\sigma_1 - 1) + (\sigma_2 - 1)}{2} = \ump(F) - 1\;.
    \end{aligned}
  \end{equation}
\end{defi}

In order to describe the bifurcation behavior of $\rpolar_{\mu,\mu_c}^\pm(F)$
as a function of the parameter $F \in \GL^+(3)$, it is helpful to
partition the parameter space $\GL^+(3)$.
\begin{defi}[Classical and non-classical domain]
To any pair of material parameters $(\mu,\mu_c)$ in the non-classical
range $\mu > \mu_c \geq 0$, we associate a \textbf{classical domain}
$\domc_{\mu,\mu_c}$ and a \textbf{non-classical domain} $\domn_{\mu,\mu_c}$.
Here,
\begin{equation}
  \begin{aligned}
    \domc_{\mu,\mu_c} &\eqdef \setdef{F \in \GL^+(3)}{\smp(\widetilde{F}_{\mu,\mu_c}) \leq 0}\;,
    \quad\text{and}\quad\\
    \domn_{\mu,\mu_c} &\eqdef \setdef{F \in \GL^+(3)}{\smp(\widetilde{F}_{\mu,\mu_c}) \geq 0}\;,
  \end{aligned}
\end{equation}
respectively.
\end{defi}

It is straight-forward to derive the following equivalent characterizations
{\small
\begin{equation}
  \begin{aligned}
    \domc_{\mu,\mu_c} = \setdef{F \in \GL^+(3)}{\ump(F) \leq \lambda_{\mu,\mu_c}}
    = \setdef{F \in \GL^+(3)}{\sigma_1 + \sigma_2 \leq \sradmm \eqdef \frac{2\mu}{\mu - \mu_c}}\;,\\
    \domn_{\mu,\mu_c} = \setdef{F \in \GL^+(3)}{\ump(F) \geq \lambda_{\mu,\mu_c}}
    = \setdef{F \in \GL^+(3)}{\sigma_1 + \sigma_2 \geq \sradmm \eqdef \frac{2\mu}{\mu - \mu_c}}\;.
  \end{aligned}
\end{equation}
}

On the intersection
$\domc_{\mu,\mu_c} \cap \domn_{\mu,\mu_c} = \setdef{F \in \GL^+(3)}{\smp(F) = 0}$,
the minimizers $\rpolar_{\mu,\mu_c}^\pm(F)$ coincide with the polar
factor $\polar(F)$. This can be seen from the form of the optimal
relative rotations in~\proporef{propo:rhat}. More explicitly, in
dimension $n = 3$ and in the non-classical limit case
$(\mu,\mu_c) = (1,0)$, we have:
\begin{equation}
  \domc_{1,0} \eqdef \setdef{F \in \GL^+(3)}{\smp(F) \leq 0}\;,
  \quad\text{and}\quad
  \domn_{1,0} \eqdef \setdef{F \in \GL^+(3)}{\smp(F) \geq 0}\;.
\end{equation}
Since the maximal mean planar strain $\smp(F)$ is related to
strain, this indicates a particular (possibly new) type of
tension-compression asymmetry.

Towards a geometric interpretation of the energy-minimizing Cosserat
rotations $\rpolar^\pm_{1,0}(F)$ in the non-classical limit case
$(\mu,\mu_c) = (1,0)$, we reconsider the spectral decomposition of
$U = QDQ^T$ from the principal axis transformation in~\secref{sec:intro}.
Let us denote the columns of $Q \in \SO(3)$ by $q_i \in \S^2$, $i = 1,2,3$.
Then $q_1$ and $q_2$ are orthonormal eigenvectors of $U$ which correspond
to the largest two singular values $\sigma_1$ and $\sigma_2$ of $F \in \GL^+(3)$. More generally, we introduce the following concept
\begin{defi}[Plane of maximal stretch]
  \label{defi:pms}
  The \textbf{plane of maximal
    stretch} is the linear subspace
  $$\mathrm{P}^{\rm ms}(F) \quad\eqdef\quad \vspan{q_1,q_2} \;\subset\; T_x\Omega$$
  spanned by the two maximal eigenvectors $q_1,q_2$ of $U$, i.e.,
  the eigenvectors associated to the two largest singular
  values $\sigma_1 > \sigma_2 > \sigma_3 > 0$
  of the deformation gradient $F \in \GL^+(3)$.
\end{defi}
In an analogous fashion, we obtain the plane of maximal stretch in the
deformed configuration as
\begin{equation}
  \mathrm{P}_{\rm def}^{\rm ms}(F)
  \quad\eqdef\quad
  \polar(F) \, \mathrm{P}^{\rm ms}(F)
  \;\subset\;
  T_{\varphi(x)}\Omega_{\rm def}\;.
\end{equation}
Previously, in our~\proporef{propo:rhat}, we have determined the
energy-minimizing relative rotations
\begin{align*}
  \widehat{R}_{1,0}^\pm(D) \eqdef
  \argmin{\widehat{R}\,\in\,\SO(3)}{\widehat{W}_{1,0}(\widehat{R}\;,D)}
  \eqdef
  \argmin{\widehat{R}\,\in\,\SO(3)}{\hsnorm{\sym(\widehat{R}D) - \id}^2}\;.
\end{align*}

\begin{figure}
  \begin{center}
  \begin{tikzpicture}
      \node at (0,0){\includegraphics[width=0.41\linewidth,clip=true,trim=0.75cm 3cm 0.75cm 2.5cm]{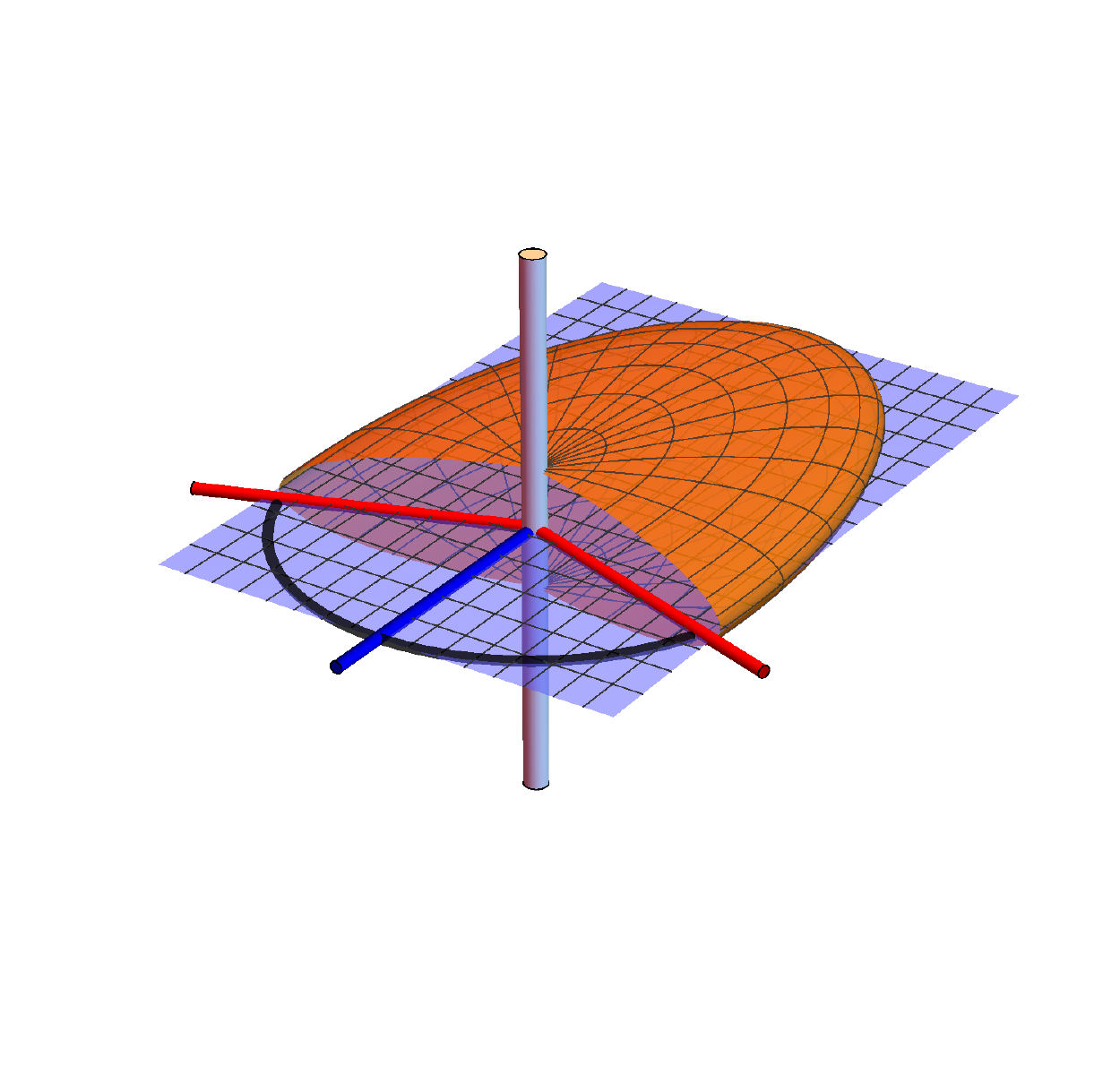}};
      \node[above] at (-2,0.5) {$-\hat{\beta}_{1,0}$};
      \draw[->] (-1.8,0.5) -- (-1, -0.25);
      \node[above] at ( 1.8, 1.3) {$\hat{\beta}_{1,0}$};
      \draw[->] (1.8, 1.3) -- (-0.1, -0.6);
      \begin{scope}[shift={(2,-1.25)},scale=0.5]
        \draw [thick,->] (0,0) -- (0.95 * -0.975, 0.95 * -0.7);
        \node[anchor=north] at (0.95 * -0.975, 0.95 * -0.7) {$q_1$};
        \draw [thick,->] (0,0) -- (1.2  * 1, 1.2 * -0.275);
        \node[anchor=north west] at (1.0  * 1, 1.0 * -0.275) {$q_2$};
        \draw [thick,->] (0,0) -- (0.025, 1.2);
        \node[anchor=south] at (0.025, 1.2) {$q_3$};
      \end{scope}
  \end{tikzpicture}
  \end{center}
  \caption{\label{fig:pms}Action of $\rpolar^\pm_{1,0}(F)$ in axes
  of principal stretch for a stretch ellipsoid with half-axes
  $(\nu_1,\nu_2,\nu_3) = (4,2,1/2)$. The plane of maximal stretch
  $\mathrm{P}^{\rm ms}(F)$ is depicted in blue. The cylinder along
  $q_3 \perp \mathrm{P}^{\rm ms}(F)$ illustrates that the axis
  of rotation is the eigenvector $q_3$ of $U$ associated with the
  smallest singular value $\nu_3 = 1/2$ of $F$. The thin cylinder [blue]
  bisecting the opening represents the relative rotation angle
  $\hat{\beta} = 0$ and corresponds to $\polar(F)$. The outer two
  cylinders [red] correspond to the  two non-classical minimizers
  $\rpolar_{1,0}^\pm(F)$. The enclosed angles
  $\hat{\beta}_{1,0}^\pm = \pm\arccos(\frac{2}{\nu_1 + \nu_2})$ are the
  optimal relative rotation angles. This reveals the major symmetry of
  the non-classical minimizers.}
\end{figure}
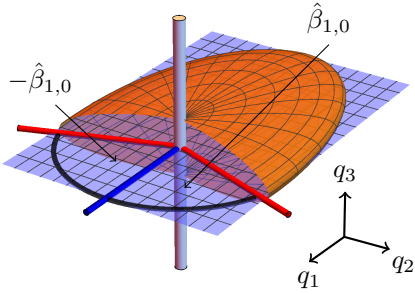

Before we proceed, it seems worthwhile to repeat that for
$\ump(F) \leq 1$, we have $F \in \domc_{1,0}$ and we have
$\widehat{R}_{1,0}^\pm(D) = \id$.
Here, the identity matrix considered as a relative rotation
corresponds to the polar factor $\polar(F)$ as an absolute
rotation. In strong contrast, the minimizers $\rpolar^\pm_{1,0}(F)$
deviate strictly from $\polar(F)$ for $\ump(F) > 1$ which
implies $F \in \domn_{1,0} \setminus \domc_{1,0}$. Only in this
latter case the relaxed-polar mechanism induces non-classical
rotation patterns.

Let us now look at the interesting non-classical case $F \in \domn_{1,0}$
in some more detail.
\begin{rem}[$\rpolar^\pm_{1,0}(F)$ in the non-classical domain]
  \label{rem:rpolar_nonclass}
  If $F \in \domn_{1,0}$, then by definition $\smp(F) > 0$
  and the maximal mean planar strain is expansive. The deviation of
  the non-classical energy-minimizing rotations $\rpolar^\pm_{1,0}(F)$
  from the polar factor $\polar(F)$ is measured by a rotation
  in the plane of maximal stretch $\mathrm{P}^{\rm ms}(F)$ given
  by $\polar(F)^T \, \rpolar^{\pm}_{1,0}(F) = Q(F)\widehat{R}_{1,0}^\mp(F)Q(F)^T$. The rotation axis is the eigenvector $q_3$ associated with
  the smallest singular value $\sigma_3 > 0$ of the deformation gradient
  $F \eqdef \nabla\varphi$ and the relative rotation angle is given by
  $\hat{\beta}_{1,0}^\mp(F) = \mp\arccos\left(\frac{1}{\ump(F)}\right)$.
  The absolute value of the rotation angle $\abs{\hat{\beta}_{1,0}^\pm(F)}$
  increases monotonically and we have the asymptotic limits
  $$\lim_{\ump(F) \,\to\, \infty} \hat{\beta}_{1,0}^\pm(F) \quad=\quad \pm \frac{\pi}{2}\;.$$
  In angle-axis representation, we obtain
  \begin{equation}
    \begin{aligned}
      \widehat{R}_{1,0}^\pm(F) &\quad\equiv\quad \left[\pm \arccos\left(\frac{1}{\ump(F)}\right),\, (0,\,0,\,1)\right]\,,\quad\text{and}\\
      \polar(F)^T\,\rpolar^{\pm}_{1,0}(F) &\quad\equiv\quad \left[\mp \arccos\left(\frac{1}{\ump(F)}\right),\, q_3 \right]\;.
    \end{aligned}
  \end{equation}
\end{rem}
In the last equation, we have used that
$\rpolar^{\pm}_{1,0}(F) = \polar(F) \, Q\widehat{R}_{1,0}^\mp Q^T$. Note the
flip of signs in the definition of the optimal branches
$\rpolar^\pm_{1,0}(F)$ arising from the definition in terms of the optimal
relative rotations $(\widehat{R}^\pm_{1,0})^T = \widehat{R}^\mp_{1,0}$.

This paper is now structured as follows: in~\secref{sec:plasticity}
we work out a connection between the variational formulation of a
full Cosserat model, the variationally optimal Cosserat rotations
$\rpolar_{\mu,0}^\pm(F)$, and plasticity theory. To this end, a
sequence of assumptions is made. Essentially, we assume that
the plastic distortion is small and rely on a best approximation
in the space of symmetric positive definite matrices due to
Higham~\cite{Higham:1988:NSPSD} (see Appendix~\ref{sec:appendix}).
In~\secref{sec:indent}, we present a synthetic nanoindentation
experiment in order to
illustrate the non-classical effects realized by the relaxed polar
mechanism $\rpolar^\pm_{1,0}(F)$. Nanoindentation is a setting of
considerable interest, because non-classical rotation patterns can
be experimentally observed, see, e.g.,~\cite{Zaafarani:2006:TITM}. Our
development in~\secref{sec:plasticity} seems well-adapted to the
setting of a nanoindentation experiment. The rotation patterns
are then compared to experiments due to Zaafarani
et al.~\cite{Demir:2009:IISE,Zaafarani:2006:TITM,Zaafarani:2008:ETIN,Zaafarani:2008:ODRP}. We conclude this paper with a
short summary in~\secref{sec:conclusion}. Finally, we briefly
present Highams best approximation in the space of symmetric
positive definite matrices~\cite{Higham:1988:NSPSD} and the
algorithmic implementation of $\rpolar_{1,0}^\pm(F)$ in the Appendix.
 \countres
\section{Multiplicative plasticity, small elastic distortions and
         Cosserat theory}
\label{sec:plasticity}
Our intention is to illustrate the non-classical \emph{relaxed-polar}
mechanism discovered in~\cite{Fischle:2016:OC2D}
and~\cite{Fischle:2016:OC3D}. Note that this is just one particular
aspect of a broad series of investigations on generalized plasticity
and microstructure by the second author; see,
e.g.,~\cite{Neff:2009:FSC,Neff:2007:RIM,Neff_Cosserat_plasticity05}
and~\cite{Neff:2004:GECBT} for extensive preliminary work by the
second author.

More precisely, we describe the energy-optimal rotation
mechanism realized by the two rotations $\rpolar_{\mu,\mu_c}^\pm(F)$,
and patterns that might arise by combinations thereof. It is
natural to expect combinations of the two patterns to arise in the
full model, since the Cosserat model has the structure of a diffuse
interface equation for the field of microrotations. We present this new
rotational mechanism in the setting of nanonindentation experiments
where non-classical rotation patterns can be
observed.\footnote{Following our nomenclature introduced
  in~\cite{Fischle:2016:OC2D} and \cite{Fischle:2016:OC3D}, the
term non-classical refers to rotation patterns which deviate from the
macroscopic field of continuum rotations $\polar(F)$.}
To this end, we carry out a comparison of the induced rotation
patterns in an idealized nanoindentation\footnote{By ``synthetic''
  and ``idealized'', we indicate that we do not consider (numerical)
  solutions of a realistic Cosserat boundary value problem.
  Towards a better understanding of the relaxed-polar mechanism it
  seems more instructive to start with the simpler case of a prescribed
  deformation mapping.}
with experimentally obtained measurements from nanoindentation into
copper single crystals. This comparison is the subject
of~\secref{sec:indent}.

\subsection{The strain energy density in isotropic
  multiplicative plasticity for negligible elastic strains}

Before we proceed to the beforementioned comparison in the next section,
we want to work out a \emph{possible} connection between the plasticity
in the deformation of crystalline metals through a series of modelling
assumptions and subsequent simplifications.
Based on this development, we show that it is possible to recover
the minimization problem with zero Cosserat couple modulus, i.e.,
with $\mu_c = 0$. This leads to the optimal Cosserat rotations
$\rpolar^\pm_{\mu,0}(F)$ which are locally energy-optimal
rotations, i.e., at every point in the domain $\Omega$.
Note that this establishes a link between the local minimization
of the shear-stretch energy with respect to rotations, i.e.,
our~\probref{intro:prob_wmm}, and the framework of
finite multiplicative plasticity theory for small internal length
scales $\Lc \ll 1$.

In what follows, we consider the setting of a nanoindentation
experiment after the removal of the indenter and the subsequent
elastic relaxation of the indented specimen.

The starting point of our development is the (by now) well-known
multiplicative decomposition
\begin{equation}
  F = \Fe\,\Fp, \quad \text{with} \quad (\Fe, \Fp)\; \in\; \GL^+(3) \times \SL(3)\;,
  \label{eq:plasticity:split}
\end{equation}
of the (macroscopic) deformation gradient
$F \eqdef \nabla\varphi$; a succinct historic account is given
in~\cite{Sadik:2015:OMD}. Note that the elastic part
$\Fe \in \GL^+(3)$ and the plastic part $\Fp \in \SL(3)$ are in
general incompatible (non-integrable): the elastic and plastic
two-point tensor fields $\Fe$ and $\Fp$ do, in general, \emph{not}
arise as the derivative of
a suitably chosen global deformation mapping, in strong contrast to the
deformation gradient field $F = \Dphi$, which they factor.

We shall make use of the following left and right polar decompositions
\begin{align}
  \Fe \;=\; \Re\,\Ue \;=\; \Ve\,\Re\;, \; \quad\text{and}\quad \Fp \;=\; \Rp\,\Up\;.
\end{align}
By $\Re \in \SO(3)$ we denote the orthogonal factor of the polar
decomposition of the elastic part (elastic rotation) and by $\Rp \in
\SO(3)$ the orthogonal factor of the polar decomposition of the
plastic part (plastic rotation). Furthermore, $\Ve \eqdef
\sqrt{\Fe\,\Fe^T}, \Ue \eqdef \sqrt{\Fe^T\,\Fe} \in \Psym(3)$ are the
left and right elastic stretch tensors, respectively, and $\Up \eqdef
\sqrt{\Fp^T\,\Fp} \in \Psym(3)$ denotes the right plastic stretch
tensor.  Applying now the polar decomposition to the multiplicative
split~\eqref{eq:plasticity:split}, we obtain
\begin{equation}
  F = \Fe\,\Fp
  = \Ve\,\Re\,\Rp\,\Up
  = \Ve\,R\,\Up\;,\quad\text{where}\quad R \eqdef \Re\,\Rp\;.
  \label{eq:plasticity:split_expanded}
\end{equation}
In what follows, we shall refer to the combined rotation $R \eqdef
\Re\,\Rp$ as the \emph{microrotation} induced by $F$. We shall also
find the related expansion
\begin{equation}
  F = \Fe\,\Fp = \Re\,\Ue\,\Rp\,\Up
\end{equation}
to be useful.

Elastoplasticity combines two fundamentally different mechanical
mechanisms.  Likewise, the mechanical energy expended in plastically
deforming a solid can be partitioned into recoverable elastic work and
plastic work. The plastic work dissipates to a large part in the form
of heat. What remains of it, stored in the solid, is referred to as
the \emph{stored energy of cold work}. In crystalline solids this
energy is stored in an evolving defect structure which is
  primarily characterized by lattice dislocations~\cite{Polyani:1934:GKP,Orowan:1934:K3MG,Taylor:1934:MPDC1,Taylor:1934:MPDC2}; cf. also~\cite{Benzerga:2005:SECW}.
We refer to the associated mechanical processes in the material
as strain hardening.

Returning to our setting of nanoindentation into copper single crystals, we
\emph{postulate} (as is often done) that the orthogonal factor of the elastic
part $\Re(\Fe) \in \SO(3)$ corresponds to rotations of the atomic lattice.
It is important to realize that the microstructure need not conform to
the macroscopic deformation which can be experimentally observed. Hence,
it is reasonable to model lattice rotations by a suitable independent
degree of freedom, e.g., a microrotation field $R$. Note further that the
elastic rotations $\Re \in \SO(3)$ are usually assumed to be reversible, see,
e.g.,~\cite{Neff_Muench_initial_plasticity_proc10,Neff:2008:CBG}, whereas
the orthogonal factor of the plastic part $\Rp(\Fp) \in \SO(3)$ is
usually assumed to correspond to irreversible plastic rotations. We further
make the assumption that the stored energy content of a copper specimen can
be well described by three contributions
\begin{equation}
  W_{\rm total} = \underbrace{W_{\rm elastic}(\Ue)}_\text{elastically recoverable}
  \;+\; \underbrace{\underbrace{W_{\rm hardening}(\Up)}_{\text{linear kin. hardening}}
    \;+\; \underbrace{W_{\rm defect}(R,\Fp)}_{\text{incompatibility measure}}}_\text{total energy of cold work}\;.
\end{equation}
This hypothesis is in general agreed on for the case of metallic
materials. Since we imagine the possibility to describe essential
effects of the deformation in an isotropic setting, we consider an
isotropic elastic energy of the type
\begin{align}
  W_{\rm elastic}(\Ue)
  &= \mue\,\hsnorm{\dev{\log \Ue}}^2 + \frac{\kappae}{2}\,\tr{\log \Ue}^2\;,
\end{align}
which is geometrically nonlinear, but physically linear in $\log(\Ue)$.
Here $\mue$ and $\kappae$ are the isotropic
Lam\'e shear modulus and the isotropic bulk modulus of linear
elasticity. Energies of this type have recently been investigated
in~\cite{Neff_Osterbrink_hencky13,Neff:2016:GLS} where they have been
finally given a natural differential geometric interpretation. Similarly,
we consider for the first contribution of the energy of cold work
\begin{align}
  W_{\rm hardening}(\Up)
  &= \mumic\,\hsnorm{\dev{\log \Up}}^2 + \frac{\kappamic}{2}\,\tr{\log \Up}^2\;.
\end{align}
Here $\mumic$ is the isotropic shear hardening modulus and $\kappamic$ is
the isotropic bulk hardening modulus.

It suffices to consider the physics of the indentation process after the
unloading procedure of the indenter was completed and the subsequent
elastic relaxation has converged to an equilibrium state. It is then
natural to presume that the elastic strains which still remain in the
specimen are very small. Thus, we consider the so-called regime of
nearly rigid plasticity which amounts to
\begin{equation}
  \hsnorm{\Ue - \id} \;=\; \hsnorm{\Ve - \id} \ll 1
  \label{eq:small_elastic_strains}
\end{equation}
and $\Fp \in \SL(3)$, i.e., plastic incompressibility. The latter is
a standard assumption in finite plasticity and we have already made it
early on. Based on our assumption that the remaining residual elastic
strains in the specimen are negligible after elastic relaxation, we
discard the elastic energy contribution alltogether and are left only
with the strain hardening and dislocation-related defect contributions.
Due to $\Fp \in \SL(3)$, the strain hardening energy simplifies to
\begin{equation}
  W_{\rm hardening}(\Up) = \mumic\,\hsnorm{\log \Up}^2\;.
  \label{eq:hardening_nonlinear}
\end{equation}
For small plastic strains $\hsnorm{\Up - \id} \ll 1$, \eqref{eq:hardening_nonlinear}
is certainly well-approximated by the quadratic energy
\begin{equation}
  W_{\rm hardening}(\Up) = \mumic\,\hsnorm{\Up - \id}^2\;,
\end{equation}
which does admit an interpretation as a measure for accumulated plastic
slip.

\subsection{Symmetric best-approximation of the plastic stretch and
            complete spatial decoupling}
It turns out that the assumption of small elastic
strains~\eqref{eq:small_elastic_strains} leads us to the following
interesting approximation of the plastic stretch $\Up$. We start
with the expansion of the deformation gradient
\begin{equation}
  F = \Ve\,\Re\,\Rp\,\Up \approx \id\,\Re\,\Rp\,\Up = R\,\Up
\end{equation}
and subsequently solve for the plastic stretch which yields
\begin{equation}
  \Up \approx \widetilde{\Up} \eqdef R^T\,F = (\Re\,\Rp)^T\,F\;.
\end{equation}
This seems fine on first glance, however, the usual geometric
interpretation of the plastic stretch tensor requires
$\Up \in \Psym(3)$
which is violated by $\widetilde{\Up}$. On these grounds a positive
definite symmetric approximation is certainly desirable.
An intuitive solution is to compute a best approximation (relative to
the Frobenius matrix norm) of $\widetilde{\Up}$ in the closed cone of
positive-semidefinite matrices which we denote by $\overline{\Psym(3)}$.
A theorem due to Higham (see~\cite{Higham:1988:NSPSD}, provided
in our Appendix~\ref{sec:appendix}) allows to precisely characterize the
unique best approximation and we shall denote it by
$\pi(\widetilde{\Up})$.
The following characterization is then intuitive: the best-approximation
$\pi(\widetilde{\Up})$ is the projection onto the symmetric part, i.e.,
\begin{equation}
  \Up \approx \pi(\widetilde{\Up}) = \sym(\widetilde{\Up})
  = \sym(R^T\,F) = \sym((\Re\,\Rp)^T\,F)\;.
\end{equation}
This simple characterization is valid as long as
$\hsnorm{F - \Re\,\Rp} < 1$, see~Appendix~\ref{sec:appendix}.
Furthermore, the defect in
symmetric positive-definiteness of $\widetilde{\Up}$ is necessarily
small due to our assumption $V_e \approx \id$.

In what follows, we introduce the contribution\footnote{It is
reasonable to consider this contribution as a so-called pseudopotential
of a dissipative process.} $W_{\rm defect}$
in terms of the square norm of a version of the Burgers' tensor
appropriate for the Cosserat setting we employ. Since such a
choice involves the derivative of $\Fp$, we are clearly in a
gradient-plasticity context (cf. also~\cite{Neff:2009:FSC}
and~\cite{Neff:2007:RIM}). More precisely, we model the
dislocation-related defect energy contribution by
\begin{equation}
  W_{\rm defect}(\Re, \Fp) \eqdef
  \frac{\mue\,\Lc^2}{2}\;
  \underbrace{\hsnorm{(\Re\Fp)^T \Curl(\Re\Fp)}^2}_\text{defect density}\;.
  \label{eq:defect_density}
\end{equation}
This dislocation-related defect-energy measures the joint incompatibility
of $\Re\Fp$. For an interpretation, note that the term $\Re\Fp$
approximates the deformation gradient in the context of our small
elastic strain hypothesis $\Ve \approx \id$, since
\begin{equation}
  \Dphi = F = \Ve\,\Re\,\Fp \approx \Re\Fp\;.
\end{equation}
Thus the proposed incompatibility measure~\eqref{eq:defect_density}
penalizes the extent to which $\Re\Fp$ is not an integrable tensor
field and can be interpreted as an approximation to the well-known
dislocation energy density
$\frac{\mue\, \Lc^2}{2}\;\hsnorm{\Fp^T \Curl \Fp}^2$.\footnote{The
independent coefficients of $R^T \Curl R$ and $R^T {\rm D}_x R$
are related by an invertible linear mapping, see~\cite[p.~153, Eq.~(3.9)]{Neff:2008:CBG}.}
Therefore, \eqref{eq:defect_density} fully takes into account the
remaining eigenstresses due to incompatibility.

Finally, we combine the approximation of the strain hardening
contribution with the previously derived dislocation-related energy
measure which yields the remaining total energy of cold work
\begin{align}
  W(\Re, \Fp) &\;=\;
  \mumic \hsnorm{\pi(\widetilde{\Up}) - \id)}^2
  + \frac{\mue\, \Lc^2}{2}\,\hsnorm{(\Re\Fp)^T \Curl(\Re\Fp)}^2\notag\\
  &\;=\;
  \mumic \hsnorm{\sym((\Re\,\Rp)^T\,F - \id)}^2
  + \frac{\mue\, \Lc^2}{2}\; \hsnorm{(\Re\Fp)^T \Curl(\Re\Fp)}^2\;.
  \label{eq:remaining_cold_work}
\end{align}
Inspection of \eqref{eq:remaining_cold_work} in the context of our
idealized nanoindentation shows that the energy of cold work
corresponds approximatively to the energy stored in the remaining
total shape change.

Finally, considering again that $\Up \approx \id$, we can also
approximate $\Fp \approx \Rp$ in the dislocation-related
defect-energy and the energy takes the form
\begin{align}
  W(\Re, \Fp) &=
  \mumic \hsnorm{\sym((\Re\,\Rp)^T\,F) - \id}^2
  + \frac{\mue\,\Lc^2}{2}\;\hsnorm{(\Re\Rp)^T \Curl(\Re\Rp)}^2\notag\\
  &= \mumic \hsnorm{\sym(R^T\,F - \id)}^2
  + \frac{\mue\,\Lc^2}{2}\;\hsnorm{R^T \Curl R}^2\;.
  \label{eq:sym_energy_with_curvature}
\end{align}

In this simplified setting, we further make the following
fundamental\footnote{There is a
  connection to the notion of a \emph{latent microstructure}
  (also: latent internal degrees of freedom), which was
  illuminated in the Discussion of~\cite{Fischle:2016:OC3D}.}
\begin{center}
{
  \setlength{\fboxsep}{10pt}
  \setlength{\fboxrule}{2pt}
  \fbox{\begin{minipage}{.9\columnwidth}
      \textbf{Assumption: (Perfectly localized microstructure)}
      \textit{The microrotations $R = \Re\,\Rp$ adjust themselves
        so as to instantaneously minimize the (approximated) cold
        work~\eqref{eq:sym_energy_with_curvature} for \mbox{$\Lc = 0$}.
        The microstructure is perfectly decoupled in the variational
        formulation in the sense that there is no spatial interaction
        of the microstructure field with itself at all.}
  \end{minipage}
}}
\end{center}

The physical relevance of this assumption is certainly debatable.
In fact, we consider it as an intermediate step ourselves, which seems,
however, necessary and quite fruitful. A better understanding of
the perfectly localized and decoupled limit case $\Lc = 0$
seems to lie at the root of a better qualitative understanding of
the Cosserat model on small length scales $0 < \Lc \ll 1$.

\begin{rem}[A model without elastic strains by nonlinear projection
    onto the complementing factors]
  Our derived model is heavily based on approximations arising from
  $\Ue \approx \id$ and $\Ve \approx \id$. Let us suppose that these
  assumptions are \emph{slightly} violated. Then our derived
  energy is a measure for the leading part of the deformation energy
  which is generated by elastic rotations and plastic processes
  \underline{only}. The replacement of the factors
  $V_e \mapsto \id$ and $U_e \mapsto \id$ with the identity
  naturally induces an energetic contribution which is
  independent of purely elastic strains.\footnote{This can be
    interpreted as a nonlinear projection onto the remaining factors
    of the expansion $\eqref{eq:plasticity:split_expanded}$.} From
  this point of view, we are simply focussing on the interaction of
  local rotations $R$ in the specimen with plastic deformation $\Fp$,
  or plastic strain $\Up$, equivalently. Essentially, in our
  development, we suppress the effects due to purely elastic
  distortions in order to obtain a clear view on the remaining effects.
\end{rem}

Towards the formulation of boundary value problems, we consider the
two-field minimization problem
\begin{equation}
  \inf_{(\varphi,R)} \int\nolimits_\Omega \mumic \hsnorm{\sym(R^T\,F - \id)}^2 + \frac{\mue\, \Lc^2}{2} \hsnorm{R^T \Curl R}^2\;\dV\;.
  \label{eq:variational_formulation}
\end{equation}
This is the geometrically nonlinear, physically linear Cosserat model
with zero Cosserat couple modulus $\mu_c = 0$, see
also~\cite{Neff_ZAMM05}. In order to gain more insight into the nature
of this variational problem, it is certainly a reasonable first step
to study the simplified case
\begin{equation}
  \inf_{(\varphi,R)} \int\nolimits_{\Omega} \mumic \hsnorm{\sym(R^T F - \id)}^2 \;\dV
  \;=\; \inf_{\varphi}  \int\nolimits_{\Omega} W^\mathrm{red}(\Dphi)\;\dV\;,
  \label{eq:min_int_wred}
\end{equation}
where the reduced Cosserat shear-stretch energy is given by
\begin{equation}
  W^\mathrm{red}(F)
   \;\eqdef\; \min_{R} \mumic \hsnorm{\sym(R^T F - \id)}^2
   \;=\; \mumic \, \min_{R} \wsym(R\,; F)\;.
\end{equation}
At this point, we have finally recovered the minimization problem as
stated in \probref{intro:prob_wmm} for the field of microrotations
$R$ which we have successfully studied in~\cite{Fischle:2016:OC2D}
and~\cite{Fischle:2016:OC3D}.

\begin{figure}
  \includegraphics[angle=83,origin=c,width=0.49\textwidth,clip=false,trim=3cm 0.5cm -1cm 0.5cm]{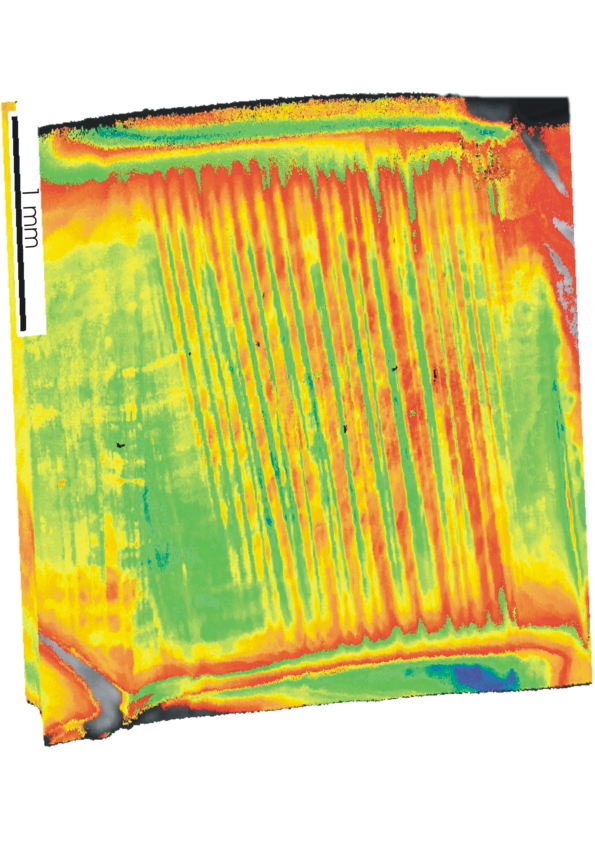}
  \raisebox{0.25cm}{\includegraphics[width=0.49\textwidth,clip=true,trim=0cm 6.5cm 0cm 7cm]{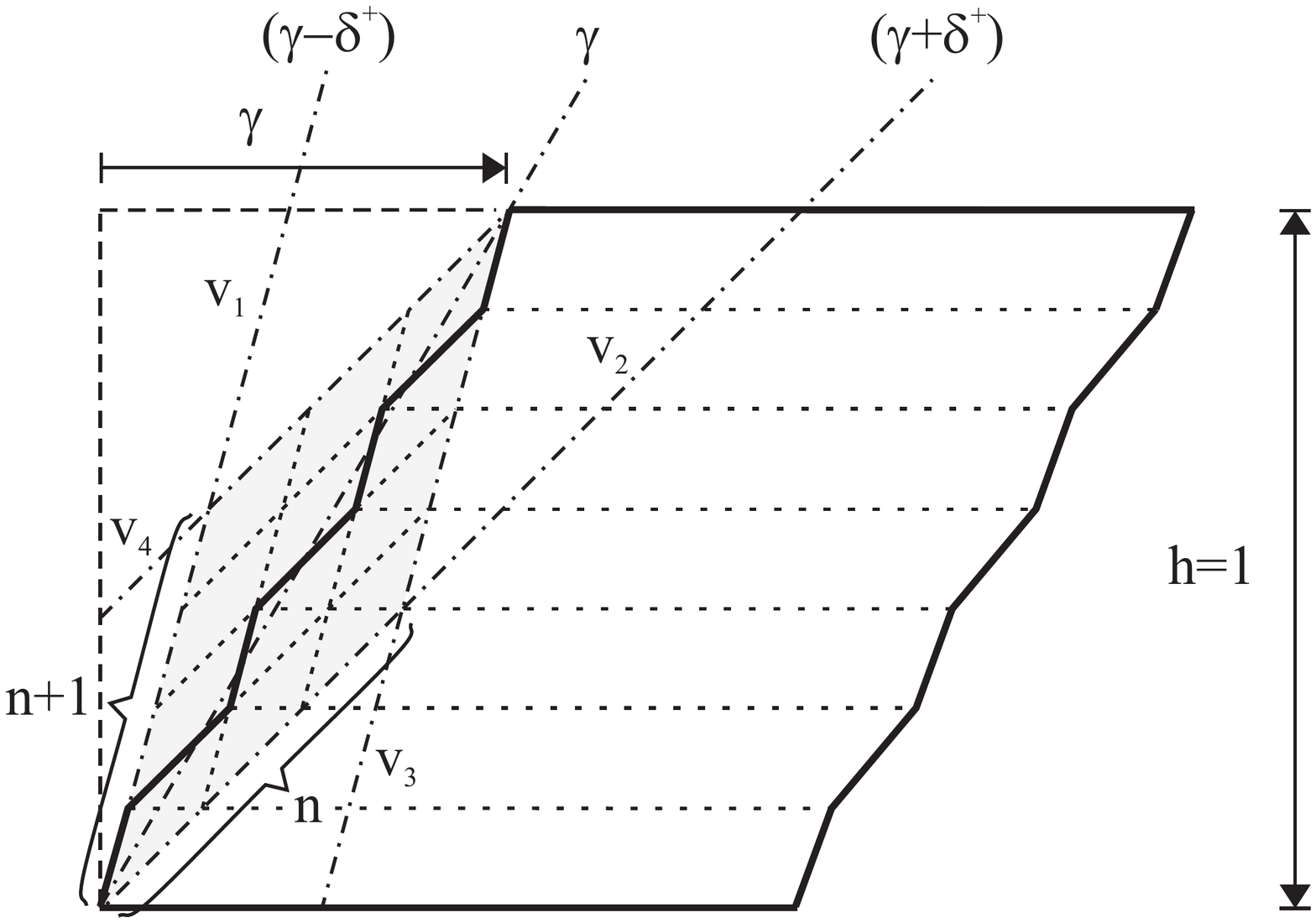}}
  \caption{\emph{Left:} a single crystal copper specimen in simple shear,
    revealing glide traces and microbands. Lattice  rotations do not
    coincide with continuum
    rotations~\cite{Dmitrieva:2009:LMS} (reproduced with permission).
    \emph{Right:} macroscopic simple shear as predicted by elastic Cosserat lamination-microstructure solution for $\mu_c = 0$; cf.~Problem~\eqref{eq:min_int_wred}.
    The homogeneous solution is a pointwise limit of the lamination
    response as the number of laminates tends to infinity. For vanishing
    characteristic length scale $\Lc = 0$ the thickness of the laminates
    remains indeterminate~\cite{Neff:2009:SSNC}.}
  \label{fig:micro_laminat}
\end{figure}

Note that the variational problem~\eqref{eq:variational_formulation}
has been studied previously in the context of simple shear of an
infinite block in~\cite{Neff:2009:SSNC}. There, an infinite strip of
height $h = 1$ is fixed at the bottom and sheared at the upper side
by an amount $\gamma \in \R$.  Solutions of~\eqref{eq:variational_formulation} are sought for in the restricted class of deformation
mappings $\varphi(x_1,x_2,x_3) = (x_1 + u(x_3),x_2,x_3)$ which are
of the form of simple glide. Similarly, microrotations
$R(x_1,x_2,x_3) = R(x_3)$ with fixed rotation axis $e_2$ are
considered. The boundary conditions
for $\varphi$ are $u(0) = 0$, $u(1) = \gamma$ and for the
microrotation field $R$ a consistency condition is imposed. It is
then possible to show that uniform simple shear is always critical,
however, non-uniform microstructure solutions exist and are
energy-optimal. The explicit solutions to the boundary value problem
realize a ``deck of cards''-mechanism which resembles real deformation
patterns also with regard to the phenomenon of counter-rotations,
see~\cite{Dmitrieva:2009:LMS} and~\figref{fig:micro_laminat}.

\subsection{Consistency of the modelling assumptions with the
            physics of a nanoindentation}

It is an important question to which degree the modelling assumptions
lead to a valid description of a physical process in a given physical
regime. In the nanoindentation experiments carried out by Raabe et al.
at the Max-Planck Institut f\"ur Eisenforschung, D\"usseldorf, which
we consider here, the total deformation is certainly to a large extent
plastic, i.e., $F \approx R\Fp$. This requires that $R\Fp$ is nearly
compatible.
(The ``synthetic'' $F$ presented in the next section is a deformation
gradient and hence compatible.)  Furthermore, the total deformation
is nearly incompressible, since it arises from translation along
glide planes. This is a common argument to justify $\Fp \in \SL(3)$.

In what follows, we consider experimental results due 3D-EBSD
measurements of nanoindentations in solid copper. The experimental
method referred here to as 3D-EBSD stands for a tomographic variant of
the electron backscatter diffraction technique that allows to
reconstruct three-dimensional crystallographic multigrain and
multiphase objects by sequential serial sectioning. The full method
was described
in~\cite{Zaefferer:2008:3DOM,Konijnenberg:2015:AGND,Konrad:2006:IOGL}.
Since the measurements in~\cite{Zaafarani:2006:TITM} were made after
removal of the indenter, the unloaded sample is elastically relaxed
but contains remaining local eigenstresses due to incompatibility.
Hence, it seems reasonable to expect that $\Ve \approx \Ue \approx \id$
is valid for the nanoindentation experiments of interest.

On the nanometer length scale, a single crystal copper sample is a
highly anisotropic medium. Nonetheless, we may ask the following
question: to which degree can the essential deformation modes due to
lattice effects, e.g., dislocation glide and lattice rotations, be
described by an isotropic model? And which essential features of the
macroscopic deformation can be characterized without resorting to the
fundamental crystal mechanics? It is mostly accepted that crystal
mechanics strongly influence the classical mechanical behavior, i.e.,
on all scales beyond the realm of quantum physics. However, as the
workings realized by anisotropic crystal mechanisms add up on ever
coarser scales they are expected to be averaged and homogenized.
Thus, the dominating effects on a coarser scale might (or might not)
allow for a good approximation by a less anisotropic, or even
isotropic, description on a suitably coarse length
scale.\footnote{The nanometer length scale of nanomechanics might
  be too fine as a mesoscale for such a homogenized material response
  to become dominant.}

In his doctoral dissertation~\cite{Zaafarani:2008:ETIN}, Nader
Zaafarani, who worked at the Max-Planck Institut f\"ur
Eisenforschung, D\"usseldorf, carried out
simulations~\cite{Zaafarani:2006:TITM,Zaafarani:2008:ETIN} based on
the \emph{physically based crystal plasticity model} with viscoplastic
hardening introduced in~\cite{Ma:2004:CMFCC,Ma:2006:MCPM} which
produced quite realistic results. Furthermore, in order to demonstrate
the failure of a traditional isotropic plasticity model, Zaafarani
carried out simulations of nanoindentation processes for an isotropic
$J_2$-plasticity model. The results clearly reflected the isotropic
consitutive model and, as expected, failed to predict the
experimentally observed non-classical rotation patterns below a
nanonindentation.

Given the complexity of physically based crystal plasticity models and
the challenges associated with their mathematical analysis, we pursue
the question whether the mathematical framework of isotropic
generalized continuum mechanics with additional degrees of freedom,
e.g., Cosserat theory, can - as an alternative - produce more
realistic results than $J_2$-plasticity. It is of interest to
determine whether isotropic micropolar and micromorphic models can
provide an alternative to physically based crystal plasticity models
\emph{on suitable length scales}, which are, however, yet to be
determined.

\countres
\section{Synthetic vs. experimental nanoindentation: a formal comparison}
\label{sec:indent}

In this section, we illustrate the geometry of the relaxed-polar
mechanism~\cite{Neff_Biot07,Fischle:2016:OC2D,Fischle:2016:OC3D} in the
setting of an idealized nanoindentation experiment. The deformation
gradient $F$ at a point and the material constants
$\mu$ and $\mu_c$ enter as parameters and generate a pitchfork bifurcation,
see~\cite[Fig.~3.1]{Fischle:2016:OC2D}.
One makes the crucial observation that the classical continuum rotation
$\polar(F)$ is always a critical point, but not necessarily a minimizer
for~\eqref{eq:intro:wmm}. In fact, in the non-classical regime
$F \in \domn_{\mu,\mu_c}$, $\polar(F)$ turns into a local maximum.
Only in this situation, the energy-minimizing branches
$\rpolar_{\mu,\mu_c}^{+}(F)$ and
$\rpolar_{\mu,\mu_c}^{-}(F)$ deviate from the polar decomposition
and generate \emph{non-classical} rotation patterns.

These non-classical Cosserat microrotation patterns motivated
us to carry out a comparison with nanoindentation experiments which
are known to produce non-classical rotation patterns below the indentation
profile; in particular so-called \emph{counter-rotations}. To be more
precise, we compare the lattice orientation angles measured by 3D-EBSD
analysis in~\cite{Zaafarani:2006:TITM} with the planar spin
$\alpha(\rpolar_{1,0}^{\pm}(F))$, see \deref{defi:planar_spin} below,
which is induced by the energy-minimizing Cosserat microrotation
fields for a synthetic nanoindentation. To this end, we have constructed
an explicit deformation mapping
$\varphi_{\rm nano}: \Omega \to \Omega_{\rm def}$
which reveals the non-classical branch of $\rpolar^\pm_{1,0}(F)$.

\begin{rem}[Choice of the Cosserat couple modulus]
  There are strong indications that a zero Cosserat coupling modulus
  $\mu_c = 0$ is the most interesting choice, first and foremost if
  one is interested in modeling a non-classical physical
  regime~\cite{Neff_ZAMM05}. We mostly focus our attention on this
  particular choice in what follows. For a quite extensive discussion
  of the expected effects due to the choice of a strictly positive
  Cosserat coupling modulus $\mu_c > 0$, we refer the reader
  to~\cite{Fischle:2016:OC2D}.
\end{rem}

\subsection{Nanoindentation in copper single crystals and 3D-EBSD analysis}

In~\cite{Demir:2009:IISE,Zaafarani:2006:TITM,Zaafarani:2008:ETIN,Zaafarani:2008:ODRP} Zaafarani et al. and in~\cite{Roters:2010:CLKH} Roters et al. reported on the orientation patterns below the indentation profiles
of nanoindentations in copper single crystals. The lattice orientations
in the deformed specimen were measured by
three-dimensional {\bf e}lectron {\bf b}ack{\bf s}catter
{\bf d}iffraction (3D-EBSD) analysis. Significant deviations of the
local lattice orientations from the macroscopic continuum rotation
have been observed and documented.\footnote{From a modelling
  standpoint, this suggests to model the crystal lattice locally as
  an independent field of microrotations $R \in \SO(3)$.}
For a detailed description of 3D-EBSD and the sophisticated setup
of the specific experiments considered here, we have to refer the
interested reader to~\cite{Zaafarani:2006:TITM} and the doctoral
dissertation of Zaafarani~\cite{Zaafarani:2008:ETIN}. Details
  of the focussed ion--beam-based experimental serial sectioning
  approach are given in~\cite{Zaefferer:2008:3DOM}.

A major motivation in \cite{Zaafarani:2006:TITM} was to compare
nanoindentation experiments carried out with a conical indenter
with simulation results obtained with a physically-based crystal
plasticity model introduced in~\cite{Ma:2004:CMFCC,Ma:2006:MCPM}.
This model class tries to capture the crystal nanomechanics
explicitly. Similarly, in our present exposition, we want to compare
experimental observations with the relaxed-polar mechanism
$\rpolar_{1,0}^\pm(F)$.

\begin{figure}[h]
  \begin{center}
    \begin{tikzpicture}
      \node[anchor=south west,inner sep=0] (image) at (0,0) {
        \includegraphics[width=10cm,trim=0cm 5cm 0cm 0cm,clip]{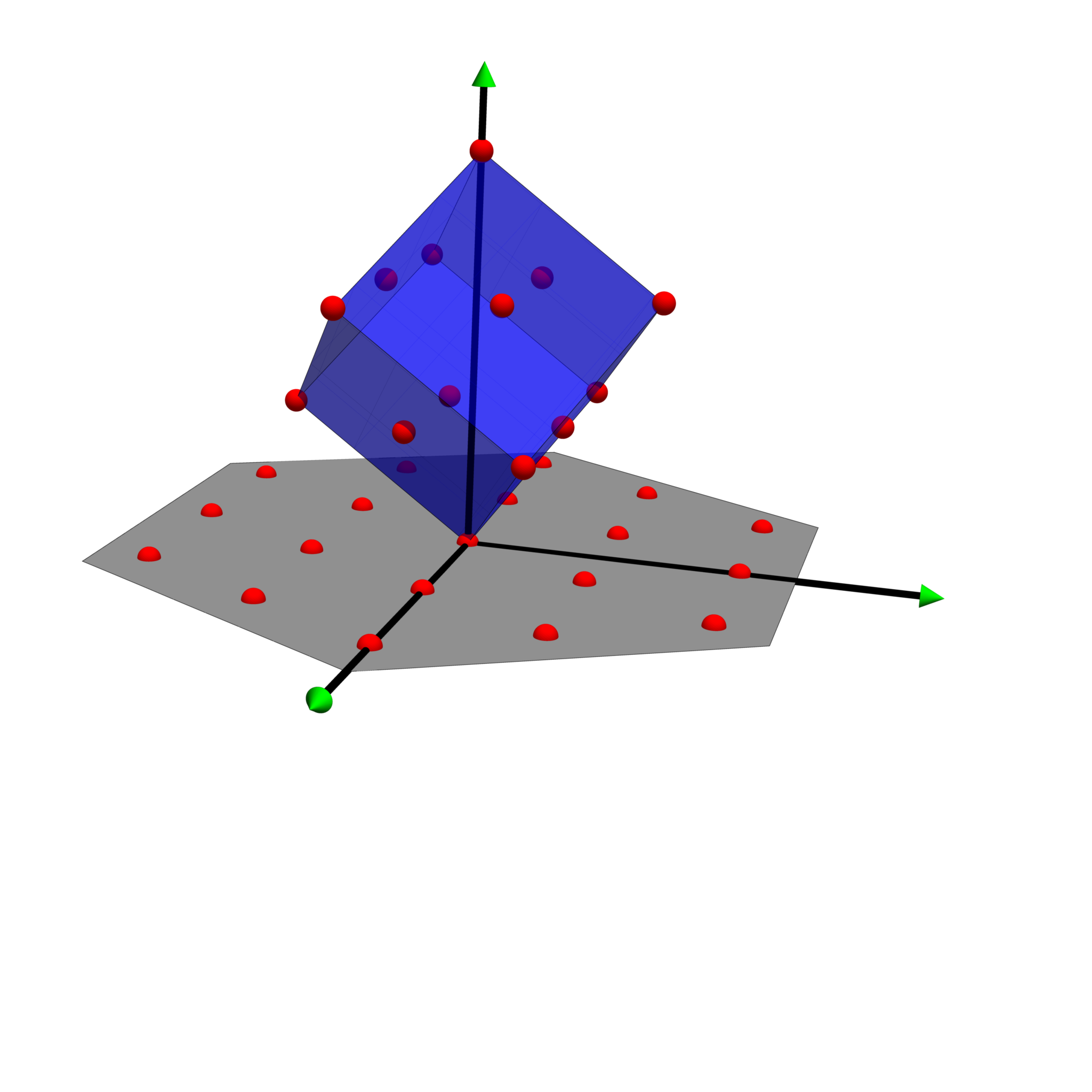}
      };
      \begin{scope}[x={(image.south east)},y={(image.north west)}]
        \node[text=black] at (0.44,1)     {$\milDir{1 1 1}$};
        \node[text=black] at (0.25,0)     {$\milDir{1 \bar{1} 0}$};
        \node[text=black] at (0.95,0.215) {$\milDir{1 1 \bar{2}}$};
      \end{scope}
    \end{tikzpicture}
  \end{center}
  \caption{Illustration of an fcc (face-centered cubic) crystal lattice made
    up of copper atoms (red), a unit cell (blue) and the $\milPlane{1 1 1}$
    crystal plane. The $\milDir{1 1 \bar{2}}$ direction is perpendicular
    to the primary slip system $\milSlipSystem{1 1 1}{1 \bar{1} 0}$.
    In~\cite{Zaafarani:2006:TITM}, Zaafarani et al. carried out nanoindentation
    experiments with indentation direction $\milDir{1 1 1}$ and EBSD section
    planes parallel to the $\milPlane{1 1 \bar{2}}$ crystal plane.}
  \label{fig:copper}
\end{figure}
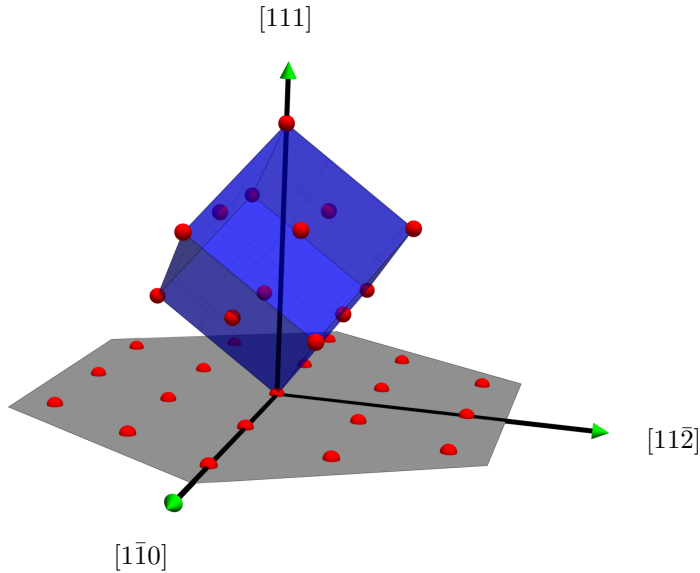

Since our exposition contains figures from~\cite{Zaafarani:2006:TITM},
we want to introduce some standard terminology from the material
sciences and some physical properties of the copper crystal lattice.
Solid copper has a face-centered (fcc) cubic crystal lattice,
as illustrated in~\figref{fig:copper}. We employ the
\emph{Miller index notation}, which is based on integer crystal
coordinates defined relative to a fixed choice of \emph{unit cell}.
By convention negative numbers are notated with a
bar $\bar{n} \eqdef -n, n \in \N$.
A unit cell of a three-dimensional crystal lattice is spanned by
\emph{direct lattice vectors}, e.g., $a_i \in \R^3$, $i = 1,2,3$,
and contains a set of representative lattice points. The lattice
points correspond to atom sites and an infinite (mathematical
abstraction of a) crystal lattice is generated by all possible
integer translations of the lattice points in the unit cell along
the direct lattice vectors $a_i$. Each lattice point can
be identified with a \emph{crystal direction} $\milDir{h k l}$.
For a cubic unit cell, e.g., a face-centered cubic copper single
crystal, the lattice vectors $a_i$, $i = 1,2,3$, span a cube.
Hence, the dual basis of (crystallographic) reciprocal lattice
vectors $a_i^*$ can be identified with the direct lattice vectors.
In this particular case, the coefficients of a \emph{crystal plane}
$\milPlane{h k l}$ in Miller index notation can be interpreted as
the coefficients of the associated normal vector $(h, k, l)$
expressed in the basis of direct lattice vectors.\footnote{In
  general, a crystal plane is defined in terms of the dual
  basis $a_i^*$.}

It is well-known that dislocation glide is a major deformation
mechanism in crystal mechanics. Dislocation glide occurs in a
\emph{slip system} (also glide system) which is a pair $(P,L)$
of a two-dimensional glide plane $\mathrm{P} \subset \R^3$ and a
one-dimensional glide direction $\mathrm{L} \subset \mathrm{P}$
contained in the plane $\mathrm{P}$.\footnote{Glide planes
  are crystal planes of closest two-dimensional atomic packings
  and glide directions are crystal directions characterized by
  closest one-dimensional packings.}
Note that in an fcc-crystal, one has, e.g., the family
of glide planes $\milPlaneFamily{1 1 1}$. The representative
glide plane $\milPlane{1 1 1}$ and the crystal direction
$\milDir{1 \bar{1} 0}$ are an example for a slip system
$\milSlipSystem{1 1 1}{1 \bar{1} 0}$.\footnote{A family of planes is
  generated by the action of the space group $\mathcal{G}_{\rm S}$
  on a representative crystal plane, here $\milPlane{1 1 1}$.
  The space group  $\mathcal{G}_{\rm S}$ is the group of affine
  linear maps which leave the lattice invariant.}

The first slip system to be activated in a given mechanical process is
referred to as the \emph{primary slip system}. In practice, multiple
slip systems can be active at the same time.  In the series of
experiments which we are interested in~\cite[Fig.~13]{Zaafarani:2006:TITM},
the pair $\milSlipSystem{1 1 1}{1 \bar{1} 0}$ is the primary slip
system. A suitable basis to study this particular system is given
by crystal directions $\milDir{1 1 1}, \milDir{1 \bar{1} 0}$
and $\milDir{1 1 \bar{2}}$; see \figref{fig:copper}.

Consider a cubic crystal lattice spanned by a right-handed system of
orthonormal direct lattice vectors $(a_1, a_2, a_3) \in \SO(3)$ in two
different orientations $R_a, R_b \in \SO(3)$. This situation arises, e.g.,
if two different grains share a common grain boundary. The following is
an intuitive
\begin{defi}[Misorientation]
  The {\bf misorientation} between the lattice orientations
  $R_a, R_b \in \SO(3)$ is given by
  $$
  R^{\rm mis}(R_a, R_b) \eqdef R_bR_a^{-1} = R_bR_a^{T} \in \SO(3)\;.
  $$
\end{defi}
Note that due to the point symmetries of the crystal lattice there are
always multiple equivalent orientations of the lattice. The minimium
over all possible misorientations is called the \emph{disorientation};
cf.~\cite[Appendix E]{Zaafarani:2008:ETIN}, or~\cite[p.~281]{Randle:2000:ITA} for an introduction.

In order to get a feeling for the spatial resolution of 3D-EBSD analysis
and also for the length scale of the experiments, we now relate the size of
a copper crystal lattice to the size of 3D-EBSD measurement points. The
current estimate for the theoretical limit of the spatial resolution
of 3D-EBSD measurements is $50 \times 50 \times 50 \pu{nm}^3$. Currently,
a realistic resolution is $100 \times 100 \times 100 \pu{nm}^3$. The
atomic radius for copper can be experimentally determined
to be approximately
$135 \pu{pm} = 1.35 \times 10^{-10} \pu{m} = 1.35 \pu{\angstrom}$.
Further, the lattice constant for a solid copper crystal at room
  temperature has been determined as
$\norm{a_i} = a = 3.597 \pm 0.004 \pu{\angstrom}$; cf.~\cite{Davey:1925:LCTM}. This defines the sidelength of a unit cell of the lattice.
  Let us divide the sidelength of a 3D-EBSD voxel by the lattice constant $a$ in order to compute the number
of crystal unit cells per side of a 3D-EBSD measurement voxel. At a
$100 \pu{nm}$ resolution, this yields
$\floor{100 \pu{nm} / 3.597 \pu{\angstrom}} = 278$
unit cells along each side of a voxel. Thus, \emph{each single measurement
  point} of the lattice orientation field in the deformed specimen contains
approximately $278^3 = 21\,484\,952$ unit cells, already at such an
extremely fine resolution. We conclude that 3D-EBSD experiments extract
misorientation mappings on a meso-scale which \emph{might}
be coarse enough to produce dominant deformation modes characterized
by meso-mechanisms. By a meso-mechanism, we mean a mechanism which
arises from the superposition of the collective deformation
and rotation behavior associated with an underlying group of
deformation carriers, i.e., dislocations in the current case,
which are generated on a smaller scale and closely related to the
specific crystal lattice.
Let us suppose that such dominant meso-mechanisms do exist; then the
consequences are two-fold: first, the question arises how closely the
meso-mechanism is still related to the crystal lattice structure which
generates it; second, since the results are homogenized, they can be
compared with simulations based on continuum mechanical models. It might
even be possible to approximate a dominant meso-mechanism
phenomenologically without introducing the fundamental crystal
mechanics which generate it. From this perspective, our interest
is to determine, whether the relaxed-polar mechanism possibly
allows to (approximatively) describe a known meso-mechanism:
the counter-rotation patterns in nanoindentation.

\subsection{Construction of a synthetic nanoindentation}

\begin{figure}[tb]
  \begin{center}
    \includegraphics[width=12cm]{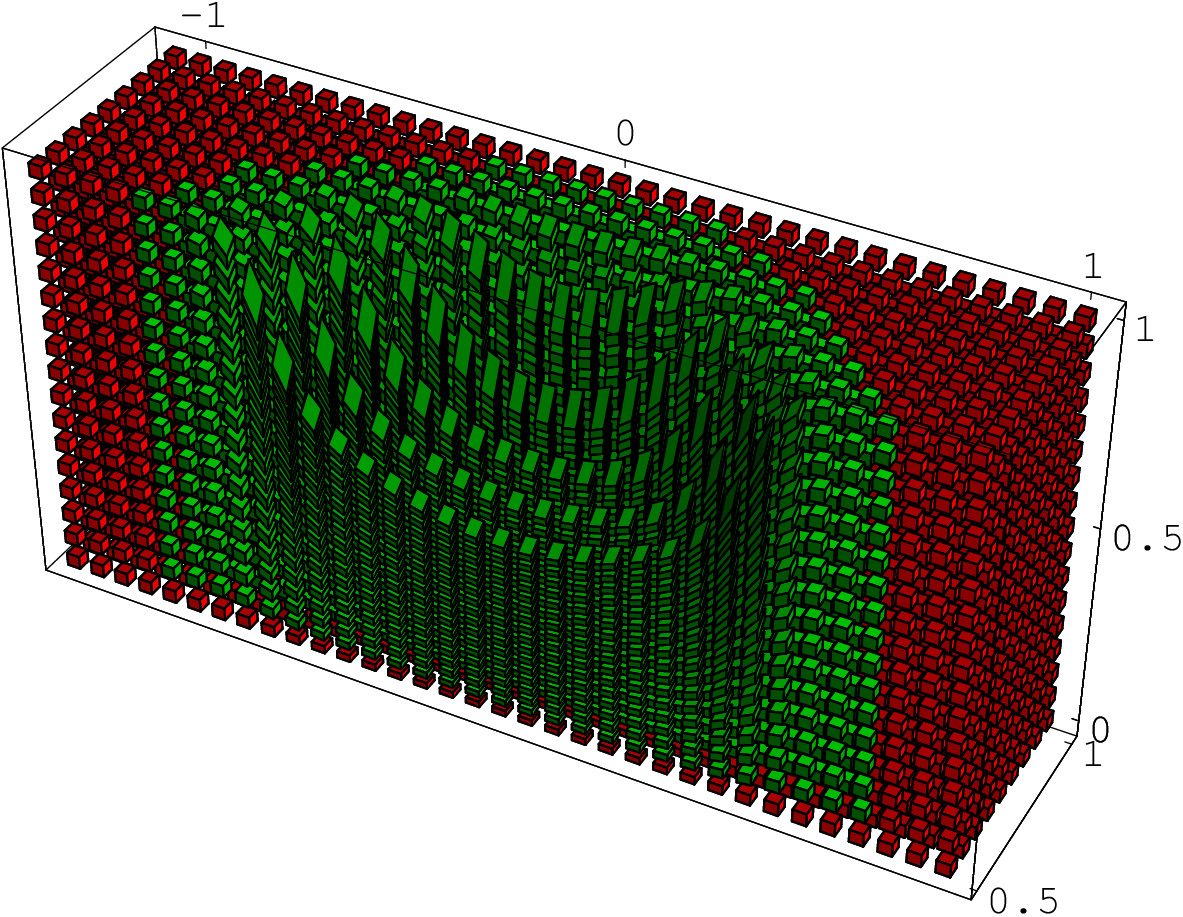}
  \end{center}
  \caption[The synthetic deformation
    $\varphi_{\text{nano}}:\Omega \to \Omega_{\text{def}}$]{Illustration of
    the synthetic deformation
    $\varphi_{\text{nano}}:\Omega \to \Omega_{\text{def}}$.
    The action of the pointwise affine linear approximation
    $\varphi_{\text{nano},\text{aff}}(p, h) = \varphi_{\text{nano}}(p) + F_{\text{nano}}(p)\,h$
    is evaluated on a grid of small cubes. Only a part
    $\setdef{(x,y,z)^T \in \Omega_\text{def}}{y \geq 0.5}$ of the deformed
    configuration is shown in order to reveal the cross-section
    $y = 1/2$. In the cylindrical inner part (green), the fields
    $\rpolar_{1,0}^{\pm}(F)$ are strictly non-classical
    and deviate from the classical continuum rotation $\polar(F)$.}
  \label{fig:indented_cube}
\end{figure}

Currently there is no experimental technique which allows
  to obtain the time-dependent deformation mapping
  $\varphi_{\rm exp}(t,\cdot): \Omega \to \Omega_{\rm def}(t)$,
  $t \in [0,T]$, from the material sample during a nanoindentation
  experiment.
  The 3D-EBSD technique measures the misorientation $R^{\rm mis}$
  of the lattice structure after the indentation process has
  completed. To understand the development of rotation patterns,
  one typically varies the applied loading in a sequence of
  indentations. Due to the sectioning process it is, however,
  currently not possible to obtain the time-dependent misorientation
  map $R^{\rm mis}(t,\cdot)$ for a \emph{single} nanoindentation
  process. Thus, there are still many open questions regarding
  the precise physical unfolding of the actual deformation process
  encoded by $\varphi_{\rm exp}(t,\cdot)$ in the crystalline specimen
  which induces the final state of the misorientation pattern
  $R^{\rm mis}(T,\cdot)$ as measured by the 3D-EBSD technique.
  In other words, it is not entirely clear, how the lattice
  misorientation patterns precisely develop in time into their
  final structure, as documented, e.g., in~\figref{fig:raabe_indent_fig13}
  and~\cite{Zaafarani:2006:TITM}.

Since the deformation mapping $\varphi_{\rm exp}(t,\cdot)$
  for a nanoindentation experiment cannot be measured yet (only
  the misorientation), we shall content ourselves with the study of
  an explicitly prescribed deformation mapping
  $\varphi_{\rm nano}: \Omega \to \Omega_{\rm def}$.
  The parametrization of $\varphi_{\rm nano}$, given below,
  models the deformation of the sample after unloading of
  the indenter at final time $T$ of an experiment and
  qualitatively matches the profile of a nanoindentation
  in~\cite{Zaafarani:2006:TITM}.
\footnote{Note that a realistic deformation mapping induced by
    a nanonindentation experiment $\varphi_{\rm exp}(t,\cdot)$
    is certainly far more complicated than $\varphi_{\rm nano}$.
    However, the reduced complexity is advantageous for the
    exposition of the relaxed-polar mechanism.}
For the reference configuration of the synthetic nanoindentation,
we choose
$\Omega = \setdef{(x,y,z)^T \in \R^3}{\norm{(x,y,z)^T}_\infty < 1}$,
which is a cube in $\R^3$ centered about the origin. This
  can be considered as a nondimensionalized sample. In the
experiments a conical indenter was used which has rotational
symmetry. This symmetry is also respected by our proposed
deformation $\varphi_{\rm nano}$.

We now list some shortcomings of our synthetic nanoindentation
$\varphi_{\rm nano}$: first of all, the deformation mapping
$\varphi_{\rm nano}$ is not a solution to a Cosserat boundary value
  problem, it is just a mapping which suitably displaces points
  vertically, i.e., from top to bottom, to create an indentation
  profile. In particular, there is no horizontal transport of
  matter. Further, we have confined the deformation
  $\varphi_{\rm nano}$ to a cylinder with radius $1$. This implies
  that the typical formation of pile-up patterns around the indentation
  profile, as investigated, e.g., in~\cite{WangRaabe04,Zambaldi:2010:PAGT,Zambaldi:2012:OINA,Zambaldi:2015:ODD}, is not accounted for by the
  synthetic nanoindentation $\varphi_{\rm nano}$.
\footnote{We expect that pile-up patterns are produced by the
  solution of the \emph{isotropic Cosserat boundary value problem},
  i.e., the variational problem \eqref{eq:variational_formulation}
  for realistic boundary conditions. However, these pile-up
  patterns cannot be expected to correlate with a particular
  crystal structure unless one enhances the elastic energy
  and the hardening contributions to model the anisotropy.
  As a first step, we can introduce a contribution
  $\scalprod{\mathbf{C}.\log \Ue}{\log \Ue}$, where
  $\mathbf{C}$ denotes the anisotropic fourth order linear
  elasticity tensor. To model an anisotropic material response
  which is more realistic at the nanometer length scale,
  one may couple the Cosserat model to a suitable single crystal
  plasticity model.}

Let us now state the explicit parametrization of
$\varphi_\text{nano}:\Omega \to \Omega_\text{def}$
on which our subsequent computations and our formal comparison are
based. Setting $r(x,y)^2 \eqdef x^2 + y^2$, we introduce
\begin{align*}
  \varphi_\text{nano}(x,y,z)^T =
  \begin{cases}
    \left(x, y, \frac{3z}{4} \left(r^2+\frac{1}{3}\right)\right)
    &\quad,\;\text{if } 0 \leq r \leq \frac{1}{2}\\
    \left(x,y,z \, (1 + \frac{3}{4}\frac{r^2-1}{\exp(\frac{1}{1-r}+\frac{2}{1-2r}) + 1})\right)
    &\quad,\;\text{if } \frac{1}{2} < r < 1 \\
    (x,y,z)
    &\quad,\;\text{otherwise}\;.
  \end{cases}
\end{align*}
For the construction of the mapping
$\varphi_{\rm nano}: \Omega \to \Omega_{\rm def}$,
we proceed as follows:\footnote{The construction uses a
  well-known bump function for the construction of a smooth
  partition of unity.}
\begin{align}
  f(t) &\;\eqdef\; \begin{cases}
    \exp(-t) & ,\quad\text{if}\quad t > 0\\
    0        & ,\quad\text{otherwise}
  \end{cases}\quad\text{,}\\
  g(r\,;a, b) &\;\eqdef\; \frac{f(b - r)}{f(b - r) + f(r - a)}\quad,\\
  h(x, y, z)  &\;\eqdef\; z \left(1 - \frac{3}{4} (1 - r^2)\, g(r\,;\frac{1}{2}, 1)\right)\quad,\quad\text{and}\\
  \varphi_\text{nano}(x,y,z)^T
  &\;\eqdef \left(\textrm{id}_{\R^2}(x,y), h(x,y,z)\right) = \left(x, y, h(x,y,z)\right)\;.
\end{align}
Note that $\varphi_\text{nano}$ is an orientation preserving
diffeomorphism, i.e., the deformation gradient field
$F_{\rm nano}: \Omega \to \Omega_{\rm def}$ takes values in $\GL^+(3)$.

The construction is rotationally symmetrical. Introducing cylindrical
coordinates $(r, \vartheta, h)$ with respect to both, the reference
and the deformed configuration, the deformation gradient is easily
seen to have the following form
\begin{equation}
  F_{\rm nano}(r, \vartheta, h)
  = {\rm D}_{(r, \vartheta, h)}\, \varphi_{\rm nano}(r,\vartheta,h)
  =
  \begin{pmatrix}
    1 & 0 & 0\\
    0 & 1 & 0\\
    (\partial_r\, \varphi_{\rm nano})(r, h)
    & 0 & (\partial_h\, \varphi_{\rm nano})(r, h)
  \end{pmatrix}\;.
\end{equation}
In particular, since $\partial_\vartheta\, \varphi_{\rm nano} = 0$ the
deformation gradient is independent of $\vartheta$. The cylindrical
symmetry in radial cross sections with constant values of $\vartheta$
allows us to visualize some properties of the synthetic indentation
$\varphi_{\rm nano}$ in the plane $\vartheta = 0$ (which coincides
with the cartesian $xz$-plane). This radial symmetry is also reflected
in the fields of relaxed polar factors $\rpolar_{1,0}^\pm(F_{\rm nano})$
and the continuum rotation $\polar(F_{\rm nano})$.

\begin{figure}[tb]
  \begin{center}
    \subfigure[Deformation of layers parallel to the $xz$-plane.]{
      \includegraphics[width=12cm]{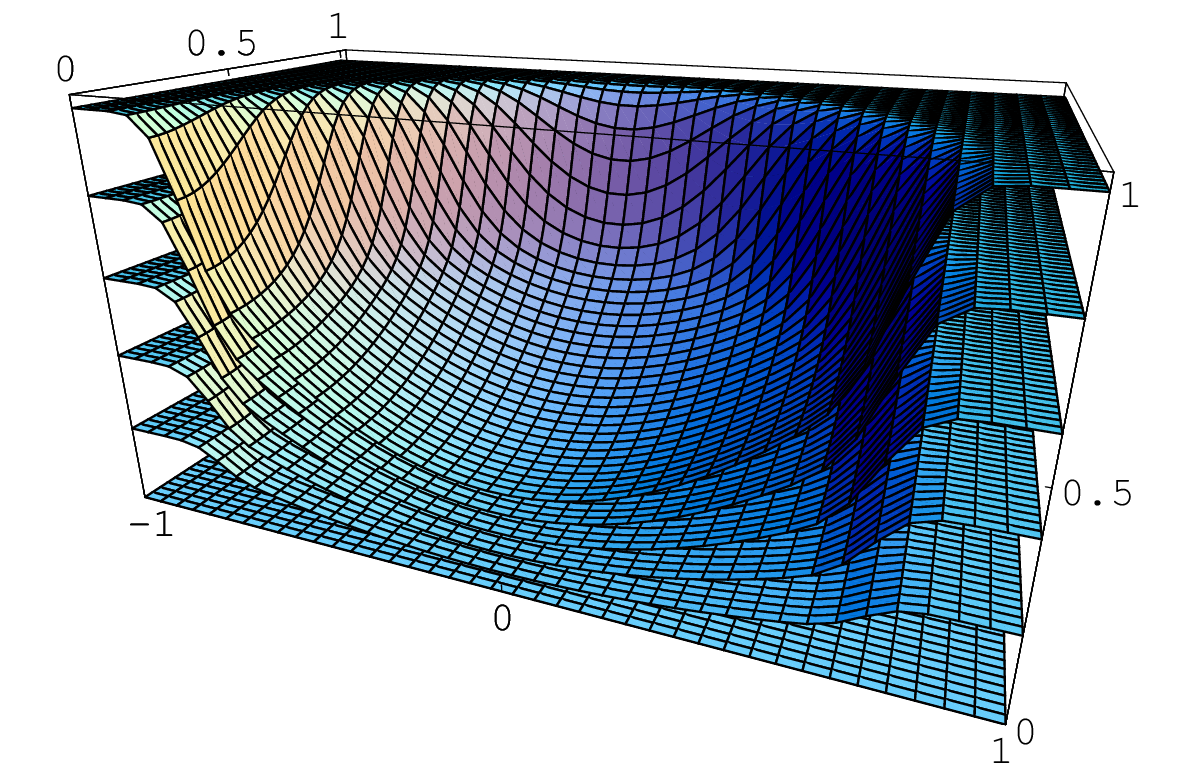}
    }
    \subfigure[Top view of indentation.]{
      \includegraphics[width=6cm]{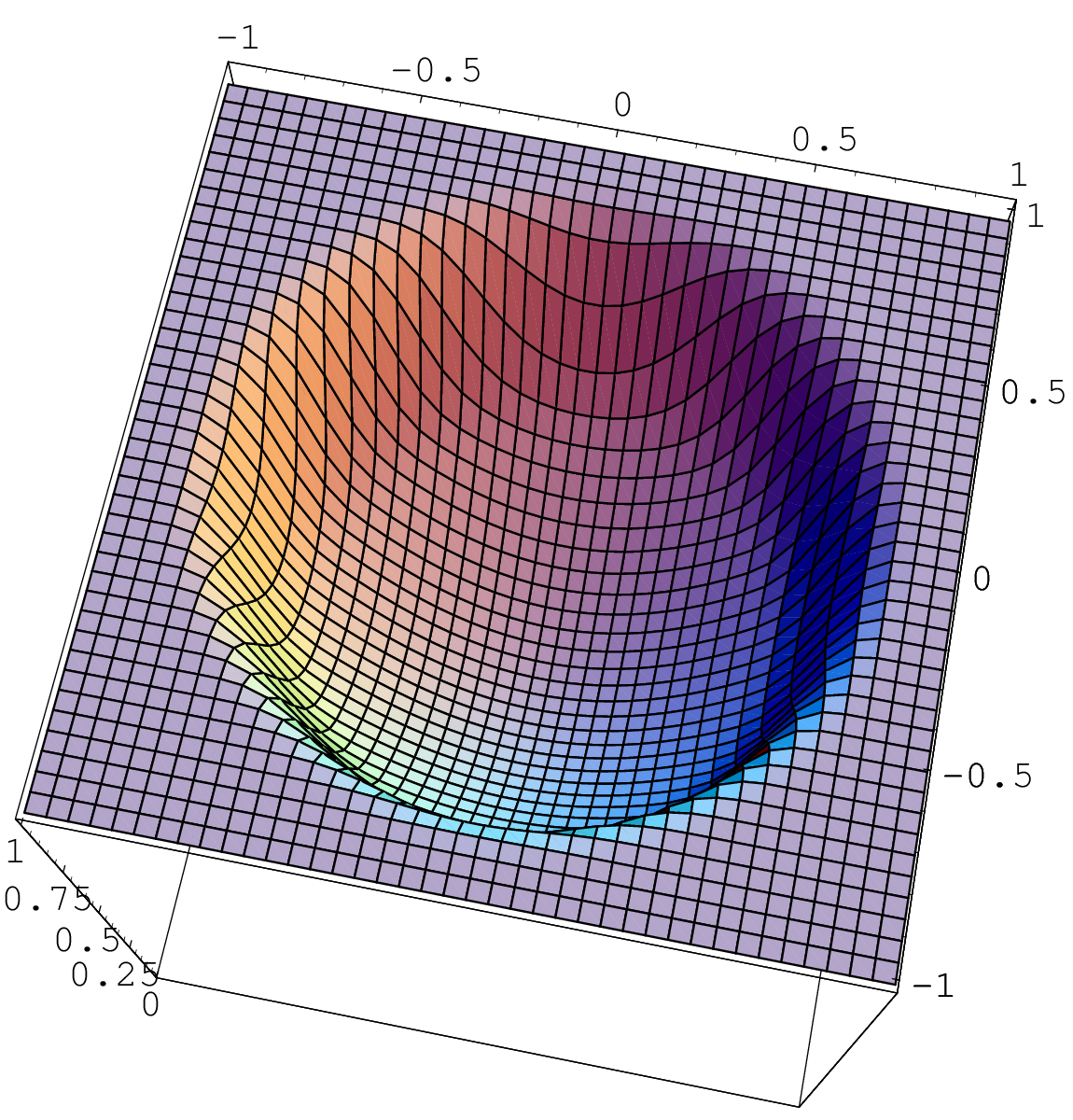}
    }
    \subfigure[Layers in the section plane {$y = 1/2$}.]{
      \includegraphics[width=6cm]{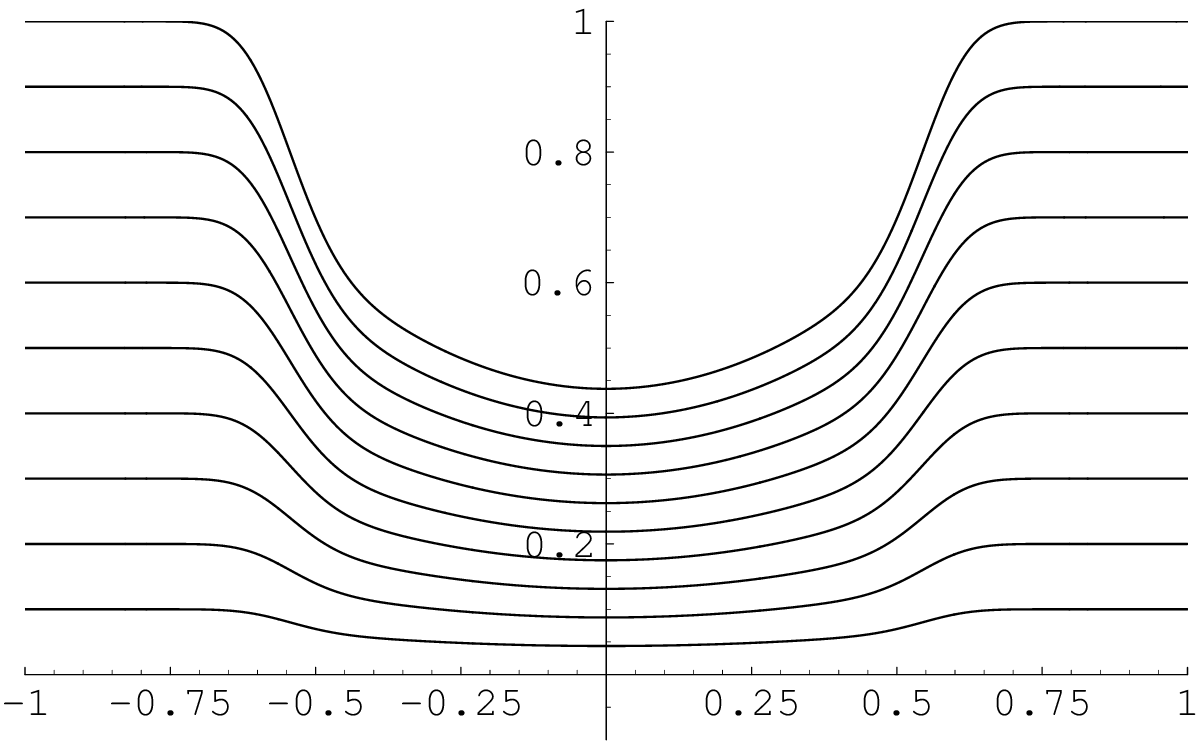}
    }
  \end{center}
  \caption[Multiple views of the synthetic deformation
    $\varphi_\text{nano}$]{Visualization of the parametrization
      of the indentation profile modelled by $\varphi_\text{nano}$.
      The coloring has no quantitative meaning and the units
      are not specified since the construction is nondimensionalized.}
  \label{fig:phi_nano_layers}
\end{figure}

\begin{figure}[tb]
  \begin{center}
    \includegraphics[clip=true,width=7cm,trim=1cm 0 6cm 3cm]{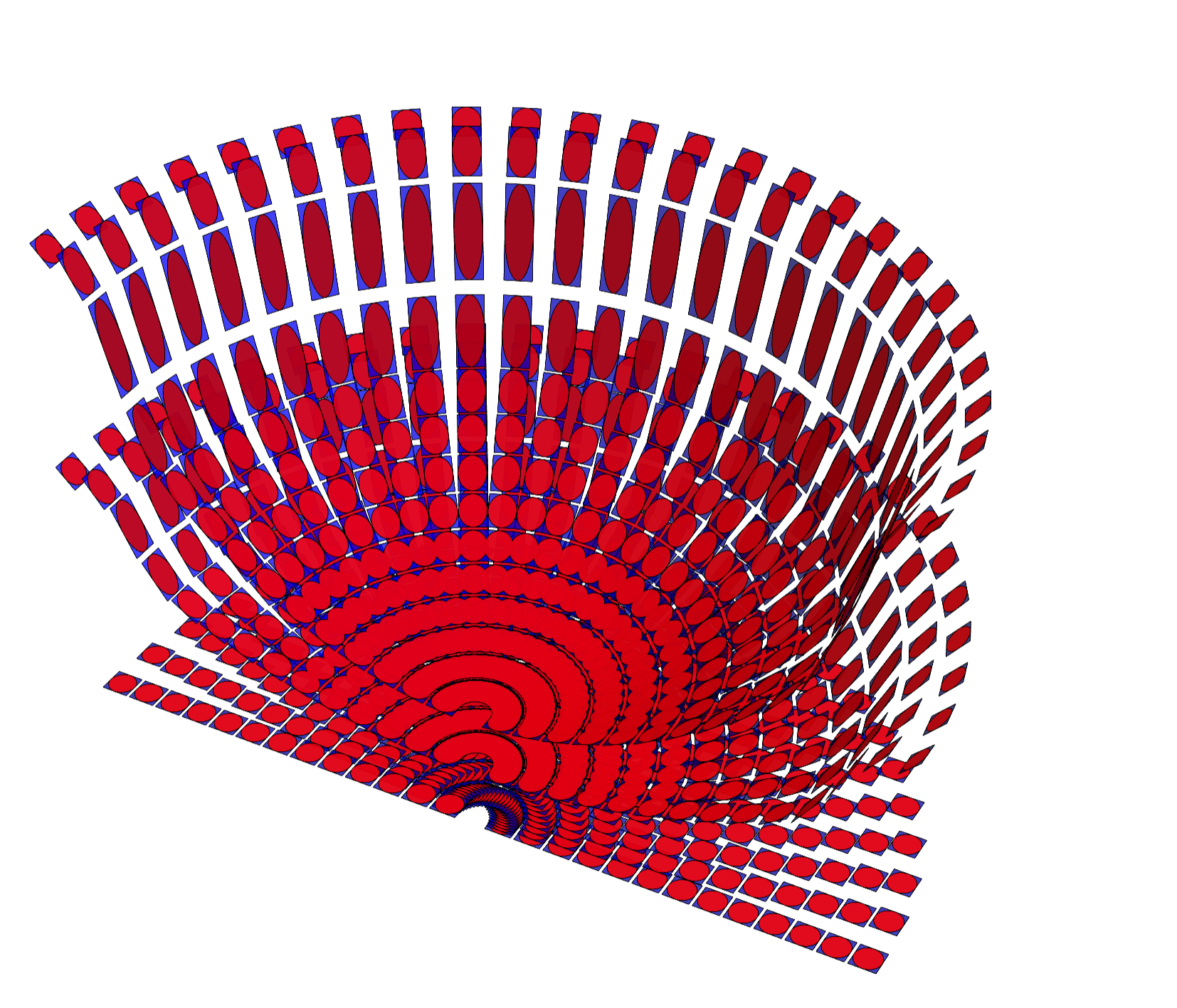}
    \includegraphics[clip=true,width=7cm,trim=1cm 0 6cm 3cm]{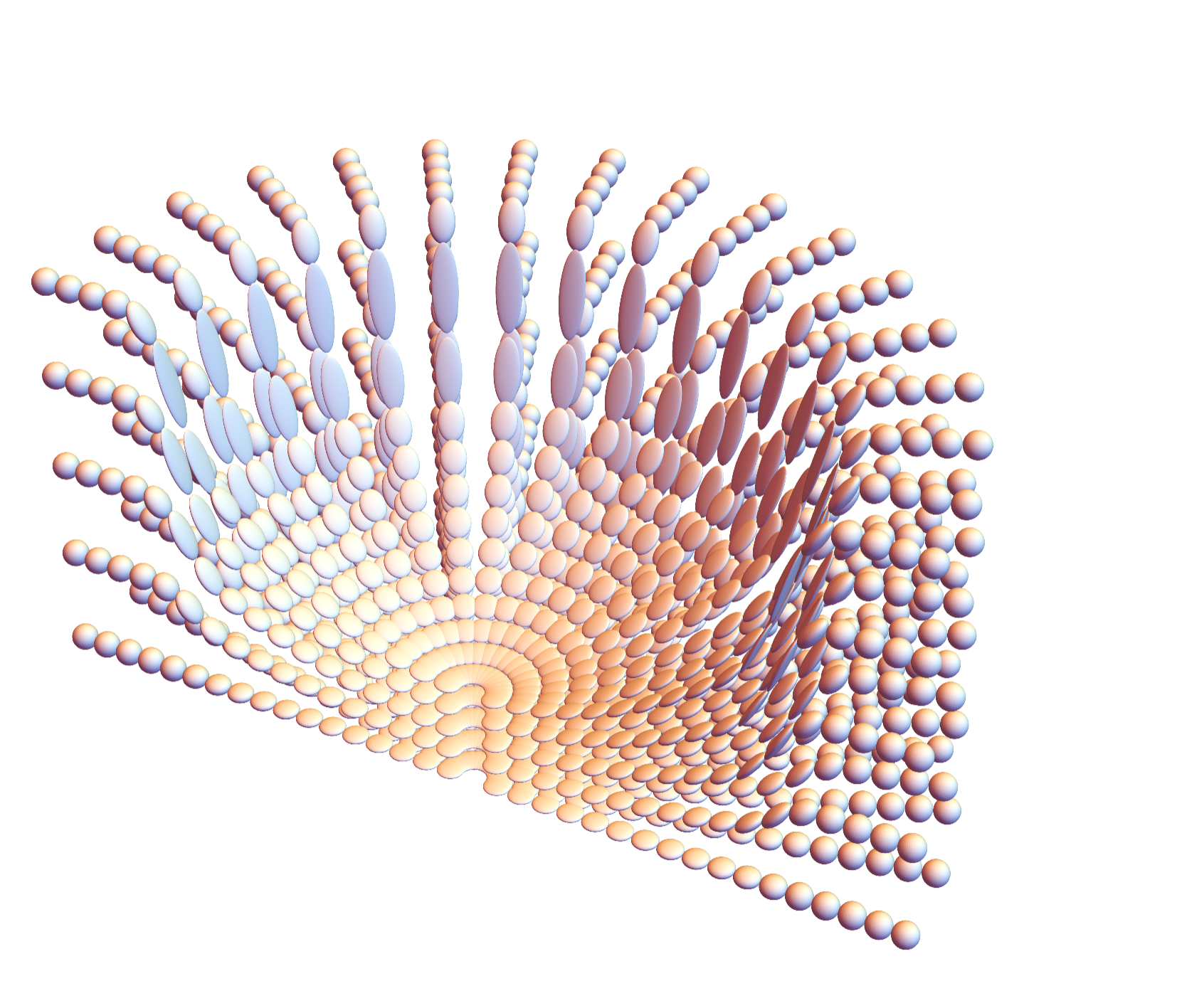}
  \end{center}
  \caption{\textit{Left:} Visualization of the plane of maximal
      stretch $P^{\rm ms}_{\rm def}(F_{\rm nano})$ [\textit{blue}]
      for the synthetic nanoindentation (in the deformed configuration)
      scaled to match the intersection with the stretch ellipsoid [\textit{red}].
      \textit{Right:} Field of stretch ellipsoids for $\varphi_{\rm nano}$.
      Note how $\polar(F) \in \SO(3)$ makes the stretch ellipsoids tumble
      symmetrically into the indentation. The coloring has no quantitative
      meaning.}
  \label{fig:plane_of_maximal_strain}
\end{figure}

In order to give the reader an impression how
$\varphi_{\rm nano}: \Omega \to \Omega_{\rm def}$ deforms the
idealized specimen $\Omega$, we have visualized the local action of
the deformation gradient field
$F_\text{nano} \eqdef \nabla\varphi_{\rm nano}$
in \figref{fig:indented_cube} and \figref{fig:plane_of_maximal_strain}.
The action of $\varphi_{\rm nano}$ on layers parallel to the
$xz$-plane is illustrated in \figref{fig:phi_nano_layers}.

\subsection{Fields of energy-minimizing Cosserat rotations \texorpdfstring{$\rpolar_{1,0}^{\pm}(F)$}{}}

In our introduction, we present the explicit form of the two
energy-equivalent optimal Cosserat rotations $\rpolar_{1,0}^{\pm}(F)$
derived in~\cite{Fischle:2016:OC3D}. The construction is pointwise
and depends on a choice of an orthonormal eigensystem
$Q \in \SO(3)$ of $U$ (or $C$, respectively). In order
to synthetically reproduce the nanoindentation experiments due
to Raabe et al. with our previously described $\varphi_{\rm nano}$,
we have to extend our definition from a pointwise one to a field
version of $\rpolar_{1,0}^{\pm}(F)$. To this end, we need
to construct an \emph{eigenframe} for $U$, i.e., a field
$Q:\Omega \to \SO(3)$ which diagonalizes $U = QDQ^T$
pointwise.

It is desirable to compute an extension which is as regular as
possible, since, in the full variational model~\eqref{eq:variational_formulation}, the gradient of the microrotation field $R$ is regularized
by the Cosserat curvature term $\hsnorm{R^T \Curl R}^2$. However, a
regular field extension, e.g., a continuous one, is more difficult to
obtain than one might think at first glance. The main
reason is that the pointwise numeric eigendecomposition of $U$ often
yields a discontinuous field of principal directions, mostly due to
jumps in eigenspace orientations.\footnote{Note that there are
  also topological constraints. To give an example, let us
  embed a smooth surface $S \subset \Omega$ which is homeomorphic
  to the two-sphere $\S^2$ and restrict $\restrict{Q}{S}$. Suppose
  that the first column $q_1: S \to TS$ is continuous. Then the
  well-known ``hairy ball theorem''
  (see, e.g., \cite[p.~382, Ex.~14-22]{JMlee02}) implies
  $q_1(p) = 0$ in at least one point $p \in S$. This contradicts
  $\norm{q_1} = 1$.}

We now state our procedure for the computation of an eigenframe in

\begin{rem}[A simple strategy for the computation of $Q$]
  Let $F \eqdef \nabla\varphi_{\rm nano}$ and suppose that
  $U \eqdef \sqrt{F^TF} \in \Sym^+(3)$ has only distinct
  eigenvalues. Furthermore, let $v_1,v_2,v_3 \in \R^3$ with
  $\hsnorm{v_i} = 1$, $i = 1,2,3$, be a numerically computed
  eigensystem of $U$. Ignoring issues of numerical stability,
  we have $V = (v_1|v_2|v_3) \in \O(3)$,
  since the eigenspaces of $U$ are pairwise orthogonal.
  For positively oriented $V \in \SO(3)$, we can set $Q = V$.
  Otherwise, we fix the orientation by reflecting the last
  column, i.e., $Q = (v_1, v_2, -v_3) \in \SO(3)$. This yields
  a positively oriented eigenframe diagonalizing $U = QDQ^T$.
\end{rem}
The fields $\rpolar_{1,0}^{\pm_{Q}}(F)$ illustrated in our various
figures have all been computed using the above strategy; cf. also
the implementation provided in Appendix~\ref{sec:appendix}.

\begin{figure}[tb]
  \begin{center}
    \includegraphics[resolution=300,width=14cm,clip=true,trim=0cm 8cm 8cm 8cm]{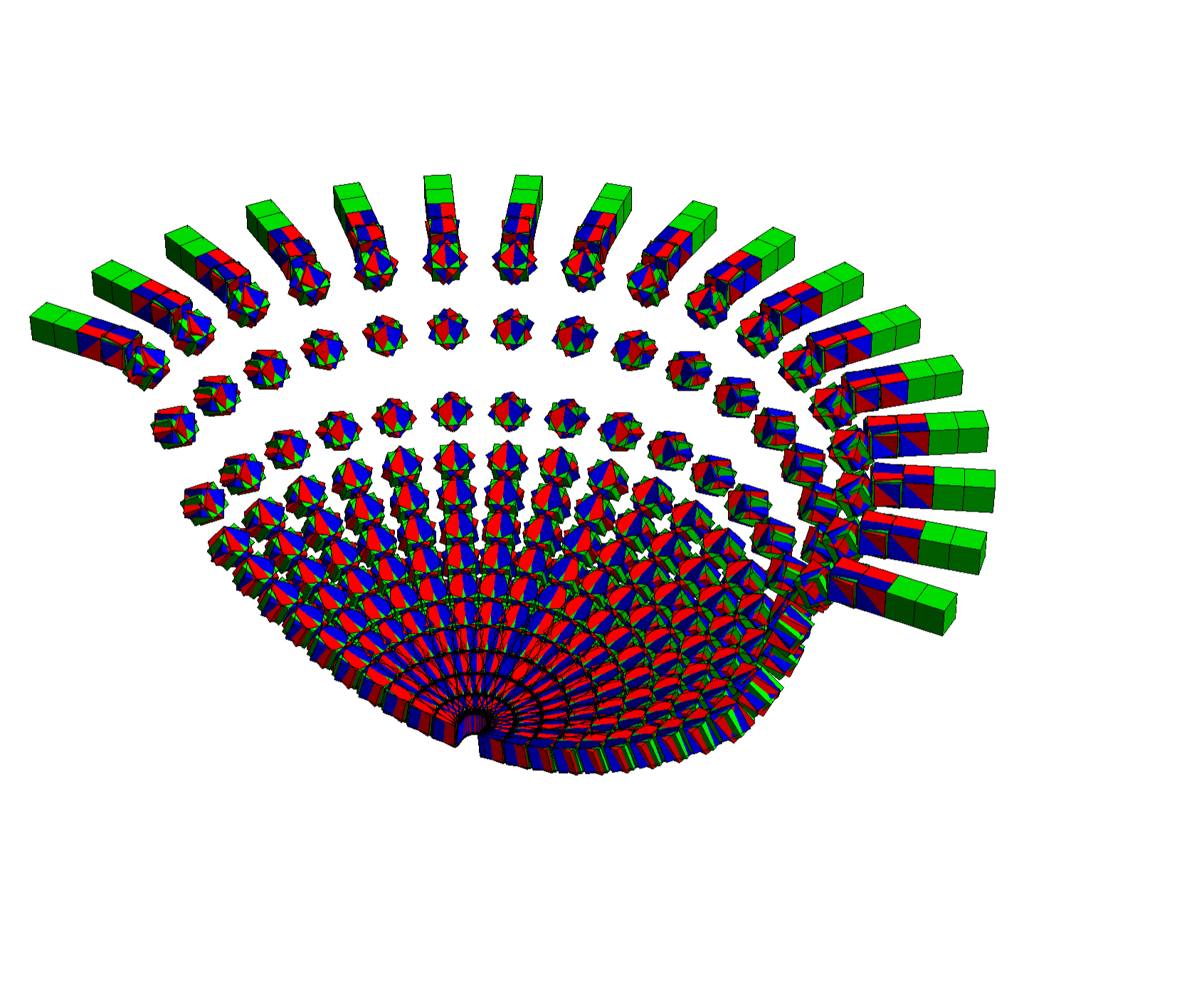}
  \end{center}
  \caption{Action of rotation fields on radially aligned
    ``infinitesimal'' cubes; the classical macroscopic continuum
    rotation $\polar(F_{\rm nano})$ [\textit{green}] tilts the cubes
    inwards into the indentation and respects the rotational
    symmetry. In strong contrast, the relaxed polar factors
    $\rpolar_{1,0}^{+}(F_{\rm nano})$ [\textit{red}] and
    $\rpolar_{1,0}^{-}(F_{\rm nano})$ [\textit{blue}] rotate the
    cubes out of the radial symmetry planes of $\varphi_{\rm nano}$.
    Note the inconsistency slightly right off the middle
    [\textit{red}/\textit{blue}]
    which illustrates a flip of the two optimal branches
    $\rpolar_{1,0}^\pm(F)$. This is due to an instability of the
    subspace orientation in the computation of the eigenframe
    $Q$ for $U$.}
  \label{fig:cube_action}
\end{figure}

\subsection{The planar spin as a measure for counter-rotations in
  a plane}
In certain experiments in structural mechanics, e.g., in micro- and
nanoindentations, but also in natural phenomena, e.g., certain types
of ground slides, one repeatedly observes non-classical rotation
patterns in the particular form of \emph{counter-rotations}. These
can take the form of local rotations in the opposite sense of
the continuum rotation $\polar(F) \in \SO(3)$. We invite the reader
to imagine a pair of counter-rotating vortices which develop in his
hot morning coffee (or tea) with a little bit of milk, excited with
a linear motion of the spoon.\footnote{There is also a notion of
  an \emph{elastic} vortex.
  Recent simulations based on the so-called ``movable cellular
  automata method''~\cite{Popov:2001:TPMCA} model internal
  friction and show interesting results in contact loadings, i.e.,
  counter-rotations, see~\cite[p.~3]{Smolin:2015:RVDC}. Note that
  there are interesting connections to Cosserat theory which might
  even be recovered as a continuum limit of the method.}

Thus we require a measure for the amount of rotation in a section
plane that is induced by the deformation
$\varphi:\Omega \to \Omega_{\rm def}$.
More precisely, we want to measure how much the deformation gradient
field $F \eqdef \nabla \varphi$ locally rotates the specimen
$\Omega$ in an oriented section plane $(\Sigma,n)$. This will also
allow us to quantify the notion of counter-rotations in a plane.
To this end, we introduce the notion of an \emph{(apparent) planar spin}
induced by a linear mapping $L \in \R^{3 \times 3}$ in a given oriented
affine section plane $(\Sigma + p, n)$, where $n \in \S^2$ is a
unit normal vector, $\Sigma = n^\perp$, and $p \in \R^3$. For the
current exposition, it suffices to assume $p = 0$. Moreover, in this
case, the oriented section plane can be identified with its
normal vector $n$.

Let $\SO(3)_n \eqdef \setdef{R \in \SO(3)}{R n = n} < \SO(3)$ be the
subgroup of rotations with fixed axis of rotation given by $\vspan{n}$
and note that the rotations in $\SO(3)_n$ leave the oriented section
plane $(\Sigma,n)$ fixed. The direction vector $n \in \S^2$ equips the
rotation axis $\vspan{n}$ with an orientation. This allows to identify
a rotation $R_n \in \SO(3)_n$ with a unique signed rotation angle
$\alpha \in (-\pi,\pi]$. As a preparation for our next definition,
  we recall the parametrization for a planar rotation
\begin{equation*}
  \check{R}: (-\pi, \pi] \to \SO(2),
    \quad
    \check{R}(\alpha) \eqdef \begin{pmatrix}
    \check{R}_{11} & \check{R}_{12}\\
    \check{R}_{21} & \check{R}_{22}
  \end{pmatrix}
  =
  \begin{pmatrix}
    \cos(\alpha) & -\sin(\alpha)\\
    \sin(\alpha) & \phantom{-}\cos(\alpha)
  \end{pmatrix} \in \SO(2)\text{.}
\end{equation*}
From this, it is easy to see that the inverse mapping is well-defined
and given by
\begin{equation*}
  \check{\alpha}: \SO(2) \to (-\pi,\pi],\quad\quad
    \check{\alpha}(\check{R}) \eqdef \sign(\check{R}_{21})\arccos(\check{R}_{11})\;.
\end{equation*}

Related notions such as the lattice disorientation angle in texture
analysis (see, e.g.,~\cite{Randle:2000:ITA}), suggest to
associate a signed rotation angle with $L \in \R^{3\times 3}$ as a means
to measure the apparent rotation due to $L$ in the oriented plane
$(\Sigma,n)$. Our present approach to compute this angle is based
on a (nonlinear) best approximation of $L$ by a rotation with fixed
rotation axis $R_n(L) \in \SO(3)_n \eqiso \SO(2)$ and with respect
to the Frobenius matrix norm. This amounts to the solution of
\begin{equation}
  \label{eq:argminLMR}
  R_n(L) \;\eqdef\; \argmin{\substack{R\,\in\,\SO(3) \\ R n\;=\;n}}{\hsnorm{L - R}^2}\text{.}
\end{equation}
In a second step, we extract the signed rotation angle from this
``closest'' rotation $R_n(L)$.\footnote{This is a distance problem in
  $\R^{3 \times 3}$ with respect to the euclidean distance measure
  $\dist_{\rm euclid}(X,Y) \eqdef \hsnorm{Y - X}$
  induced by the Frobenius matrix norm.}

Given $Q \in \SO(3)$, such that $Q e_3 = n$, we can consider the
oriented section plane defined by $(\Sigma = \vspan{q_1,q_2}, q_3)$
and find that $R \in \SO(3)_n$ can be parametrized by
\begin{equation}
  R(\alpha) = Q \begin{pmatrix} \cos(\alpha) & -\sin(\alpha) & 0\\
    \sin(\alpha) & \phantom{-}\cos(\alpha) & 0\\
    0 & 0 & 1
  \end{pmatrix}Q^T\;\text{.}
\end{equation}
Hence, the problem \eqref{eq:argminLMR} reduces to finding a closest
planar rotation from
\begin{equation}
  R_n(L) \eqdef \argmin{\substack{R\,\in\,\SO(3)\\
      R n\;=\;n}}{\hsnorm{L - R}^2} =
  \;Q\,\left\{\argmin{\check{R}\,\in\,SO(2)}{\hsnorm{Q^TLQ -
      \begin{pmatrix}
        \check{R} & 0\\
        0       & 1
    \end{pmatrix}}^2}\right\}\,Q^T\text{.}
\end{equation}
Only the upper left $2 \times 2$ block varies under different
values of the rotation $\check{R}$. Introducing
\begin{align}
  \check{L}_n(L) \eqdef
  \begin{pmatrix}
    (Q^TLQ)_{11} & (Q^TLQ)_{12}\\
    (Q^TLQ)_{21} & (Q^TLQ)_{22}
  \end{pmatrix}\;,
\end{align}
it suffices to solve the two-dimensional problem
\begin{align}
  \check{R}_n(L) \eqdef \argmin{\check{R}\,\in\,\SO(2)}{\hsnorm{\check{L}_n(L) - \check{R}}^2}\;.
  \label{eq:planar_spin_2D_reduction}
\end{align}
For the sake of completeness, we want to mention that the original
spatial rotation $R_n(L) \in \SO(3)_n$ can be uniquely recovered
from the solution $\check{R}_n(L)$ to the planar
problem~\eqref{eq:planar_spin_2D_reduction}.

Assuming that $\check{L} \in \GL^+(2)$, Grioli's theorem shows that
the unique best approximation $\check{R}_n$ with respect to the
Frobenius matrix norm is the polar factor
$\polar(\check{L}) \in \SO(2)$.
Most conveniently, an explicit expression for $\polar(\check{L})$ is
available in dimension $n = 2$. Moreover, for all $n > 1$, the best
approximation can be shown to coincide with those for the geodesic
distance on $\SO(n)$ equipped with its natural bi-invariant Riemannian
metric. In particular, we have~\cite{Lankeit:2014:MML,Neff:2014:LMP}
\begin{equation}
  \polar(\check{L})
  \;=\; \argmin{\check{R}\,\in\,\SO(2)} \dist_{\rm euclid}(\check{L}, \check{R})
  \;=\; \argmin{\check{R}\,\in\,\SO(2)} \dist_{\rm geod}(\check{L}, \check{R})\;\;.
\end{equation}
Although the closest rotation (best approximation) is the same for both the
extrinsic (euclidean) and the intrinsic (Riemannian) distance measure,
the minimal distance (approximation error) itself is
different.

The preceding development motivates
\begin{defi}[The planar spin]
  \label{defi:planar_spin}
Let $L \in \GL^+(3)$ and $\check{L}(L) \in \GL^+(2)$ as before, then
\begin{equation}
  \check{R}_n(L)
  \;\eqdef\;
  \argmin{\check{R}\,\in\,\SO(2)}{\hsnorm{\check{L}_n(L) - \check{R}}^2}\;.
\end{equation}
The \textbf{planar spin} induced by $L$ in the oriented section
plane $(\Sigma, n)$ is
\begin{equation}
  \alpha_n:\GL^+(3) \to (-\pi,\pi],\quad\quad
    \alpha_n(L) \;\eqdef\; \check{\alpha}(\check{R}_n(\check{L}_n))\;\text{.}
\end{equation}
\end{defi}
Although $Q$ is not uniquely defined in the definition of $\check{L}_n$,
the planar spin $\alpha_n(L)$ is well-defined since rotations about the
same axis $n$ always commute.

\begin{exa}
  Let $(\Sigma,n) = (\vspan{e_1,e_2}, e_3)$ and $L \in \R^{3\times 3}$.
  Further, we denote the upper left $2\times 2$ block of $L$
  by $\check{L}$. In this case, we have $n = e_3$ and $Q = \id$
  and we find
  \begin{equation*}
    R_n(L) = \begin{pmatrix}
      \check{R}_n(L) & 0\\
      0          & 1
    \end{pmatrix}\;.
  \end{equation*}
  Thus, with the explicit form of the polar decomposition in dimension
  $n = 2$ (see, e.g.,~\cite{Fischle:2007:PCM}), we obtain
  \begin{align}
    \check{R}_n(L) \;=\; \polar(\check{L})
    \;=\;
    \frac{1}{\sqrt{\norm{\check{L}}^2 + 2\;\det{\check{L}}}}
    \begin{pmatrix}
      -(\check{L}_{11} + \check{L}_{22}) & \check{L}_{12} - \check{L}_{21}\\
        \check{L}_{21} - \check{L}_{12}  & -(\check{L}_{11} + \check{L}_{22})
    \end{pmatrix}\;\text{.}
  \end{align}
\end{exa}

\begin{rem}[Approximation error of the planar spin]
  A best approximation is only guaranteed to be a good approximation
  if the associated approximation error is small. This need not be the
  case for the proposed planar spin measure $\alpha_n(L)$. Intuitively
  spoken, the measured effects coincide with our expectations if
  $L \in \R^{3 \times 3}$ essentially acts only in the section plane
  $(\Sigma, n)$. On the other extreme, if $L$ is a rotation about an
  axis $d \perp n$, i.e., $d \in \Sigma$, the planar spin degenerates.
  We may even conclude that the observation of counter-rotations in
  a section plane which does not conform well with the actual mechanical
  workings that happen during the deformation of the specimen, may in
  fact be misleading our intuition.
\end{rem}

\subsection{A formal comparison of rotation patterns in experimental
  and synthetic nanoindentation}

We are now in the position to present a formal comparison
of the non-classical rotation patterns arising in real
nanoindentation experiments below the indentation profile
which we oppose with the planar spin $\alpha(\rpolar^\pm(F_{\rm nano}))$
due to our synthetic nanoindentation $\varphi_{\rm nano}$.

\begin{figure}[tb]
  \begin{center}
    \includegraphics[width=12cm]{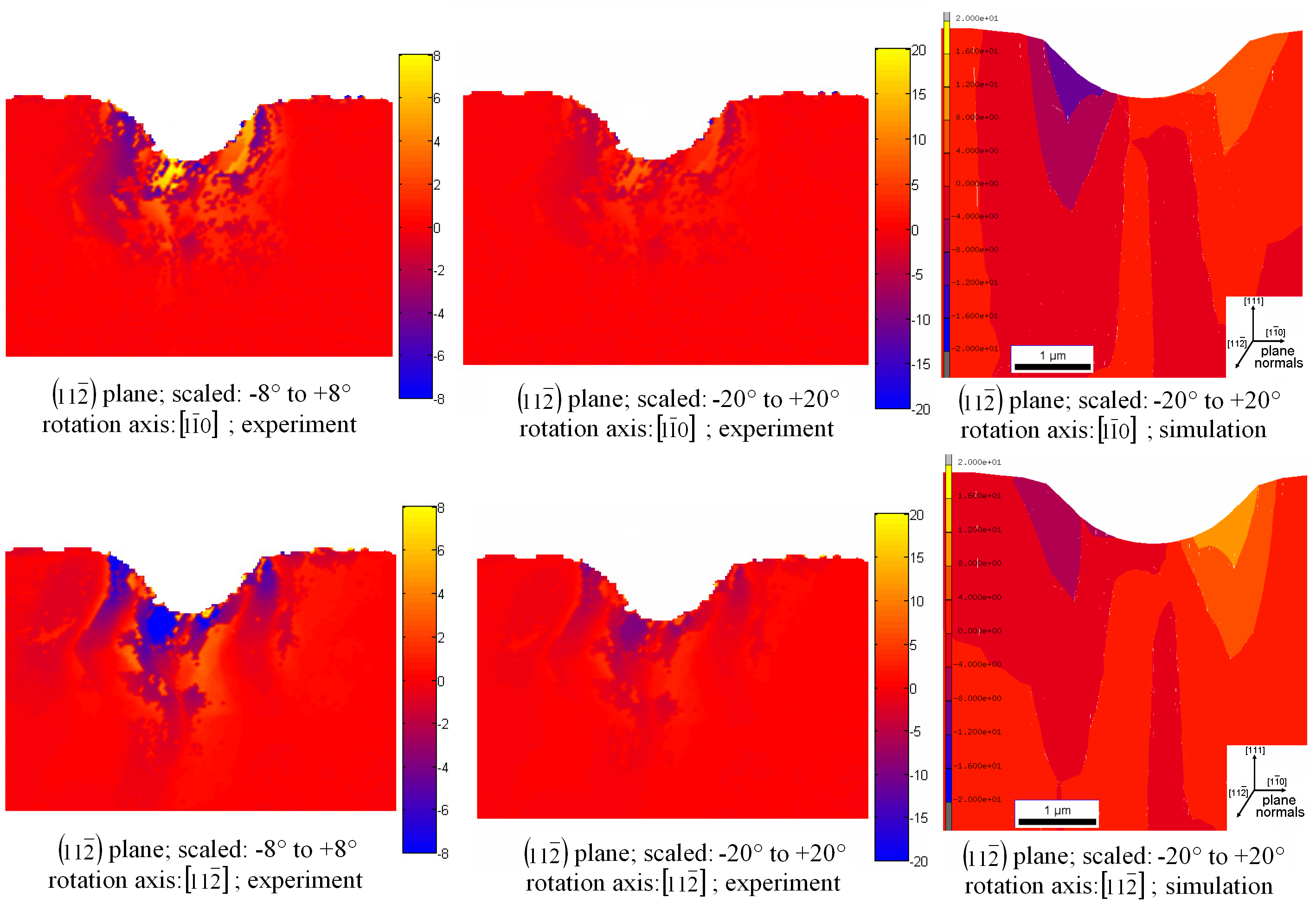}
  \end{center}
  \caption{\textit{Left:} Disorientation angle relative to the
    $\milDir{1 1 0}$ (top) and $\milDir{1 1 \bar{2}}$ (bottom)
    crystal directions. The counter rotations correspond well
    with the planar spin in the section plane $y = 1/2$ induced
    by our synthetic indentation, see~\figref{fig:planar_spin_patched}.
    \textit{Right:} Disorientation angle obtained by physics-based
    crystal plasticity simulations relative to
    the $\milDir{1 1 0}$ (\textit{top})
    and $\milDir{1 1 \bar{2}}$ (\textit{bottom}) rotation
    axes with a color scale ranging from $\pm 8^\circ$.
    (Reprint of~\cite[Fig.~13]{Zaafarani:2006:TITM}, with
     permission)}
\label{fig:raabe_indent_fig13}
\end{figure}

In \figref{fig:raabe_indent_fig13} and \figref{fig:counter_rotations_ebsd}
(courtesy of Raabe et al.), we show 3D-EBSD measurements of the
crystal lattice misorientation
after a series of nanoindentations in a copper single
crystal~\cite{Zaafarani:2006:TITM}. The figures illustrate the
experimentally obtained rotational deviation of the deformed
lattice relative to its original state. More precisely,
both, \figref{fig:raabe_indent_fig13} and
\figref{fig:counter_rotations_ebsd},
display the lattice disorientation expressed as signed rotation
angles relative to the $\milPlane{1 1 \bar{2}}$ crystal plane;
cf.~\cite[Appendix E]{Zaafarani:2008:ETIN}.
Note the typical pattern of counter-rotations separated
by multiple cross-over zones. Furthermore, comparisons
with crystal-plasticity simulations based on a model due
to \cite{Ma:2004:CMFCC,Ma:2006:MCPM} are presented in
\figref{fig:raabe_indent_fig13} which show favorable
qualitative results.

\begin{figure}[tb]
  \begin{center}
    \begin{minipage}{12cm}
      \subfigure[{$\rpolar_{1,0}^{+}(F_\text{nano})$}]{
        \includegraphics[width=6cm]{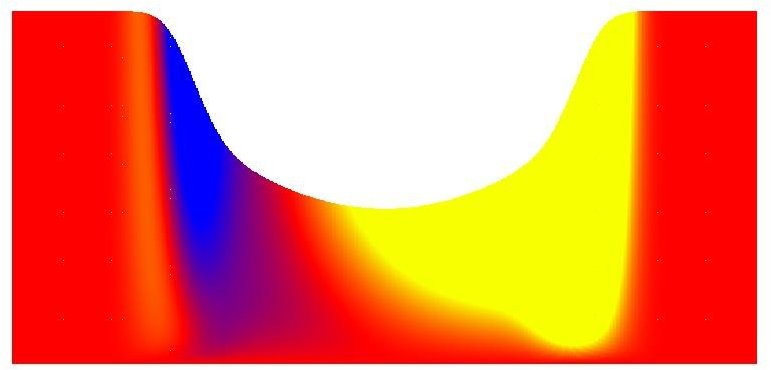}
      }
      \subfigure[{$\rpolar_{1,0}^-(F_\text{nano}($}]{
        \includegraphics[width=6cm]{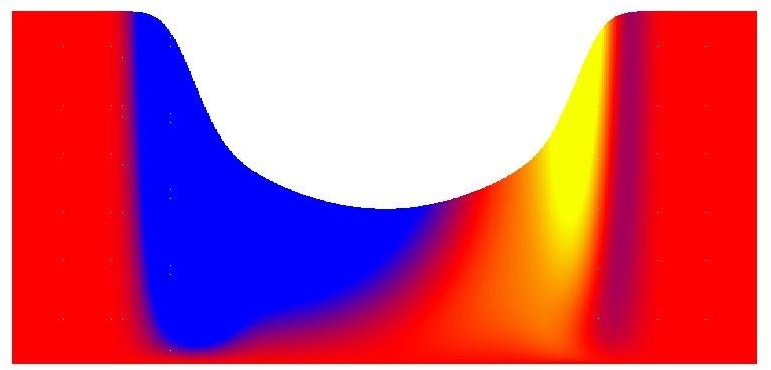}
      }\\
      \subfigure[{$\polar(F_\text{nano})$}]{
        \includegraphics[width=6cm]{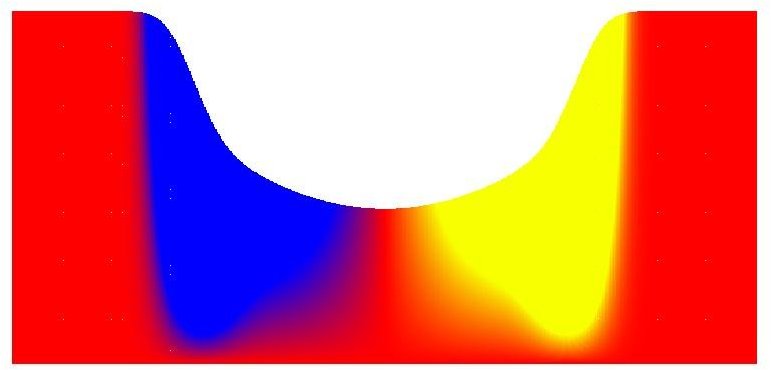}
      }
      \subfigure[{``Collage'' of $\rpolar_{1,0}^{\pm}(F_\text{nano})$}]{
        \includegraphics[width=6cm,clip=true,trim=0cm 6.8cm 0cm 4cm]{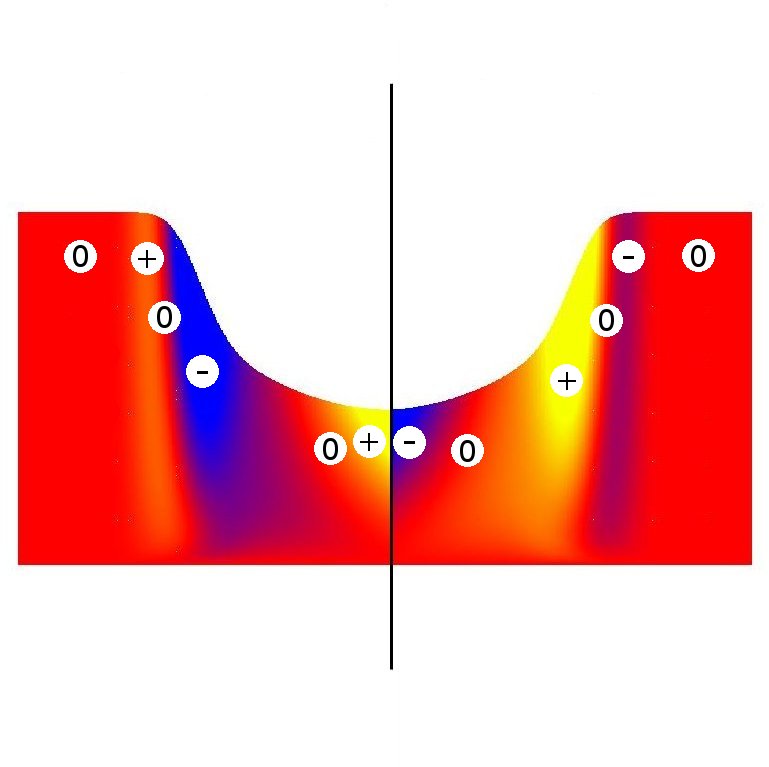}
      }
    \end{minipage}
    \hspace{0.5cm}\raisebox{-1.5cm}{\includegraphics[width=1.3cm]{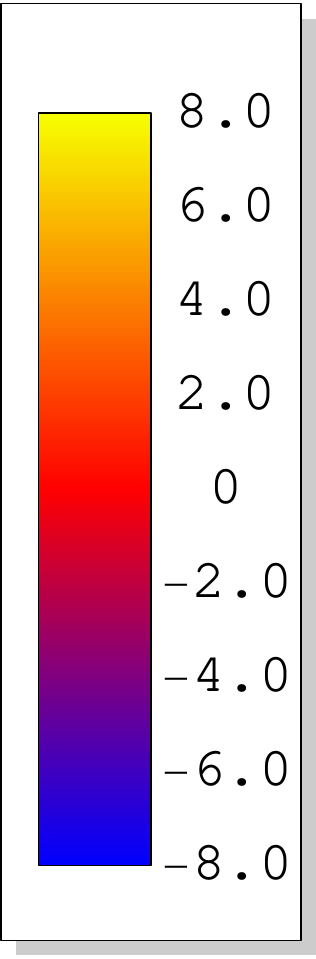}}
  \end{center}
  \caption{The planar spin $\alpha(F_{\rm nano})$ in the intersection
    plane $y = \frac{1}{2}$ of $\rpolar^\pm(F_\text{nano})$, an optimal
    choice of branches, and the continuum rotation $\polar(F_\text{nano})$
    with a color scale ranging through $\lbrack -8^{\circ},8^{\circ}\rbrack$.
    The section plane is parametrized by $(x,\frac{1}{2},z)$ and the marks
    illustrate the sign of the planar spin and the cross-over
    points. \emph{Note: the two outermost, slim zones of counter-rotations
      in the visualization may in fact be due to an instability of
      $\rpolar_{1,0}^\pm(F)$ for $F \approx \id$. The large zones of
      counter-rotations are reliable.}}
  \label{fig:planarSpins}
\end{figure}

\begin{figure}[h!]
  \begin{center}
    \subfigure{
      \includegraphics[width=8cm]{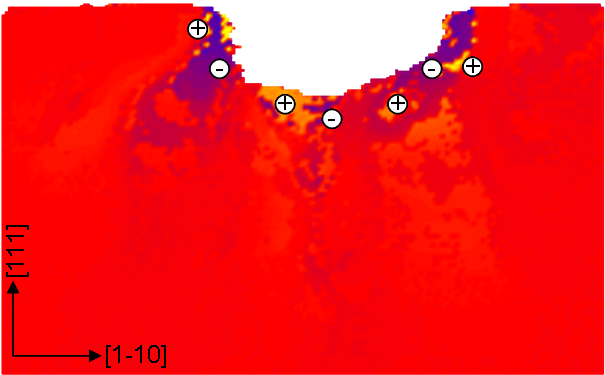}
      \raisebox{0cm}{
        \includegraphics[height=5cm]{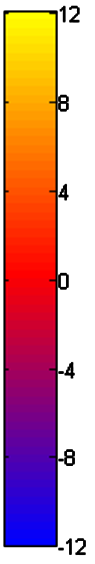}}
    }
  \end{center}
  \caption{Rotation angles measured by 3D-EBSD analysis of a nanoindentation
    into a copper single crystal with a color map scaled
    to $\pm 8^\circ$. It is important to note the cross-over zones
    separating counter-rotations. The angles are computed with respect
    to the rotation axis $\milDir{1 1 \bar{2}}$ for rotations in the
    $\milPlane{1 1 \bar{2}}$-plane (courtesy of Raabe et al., see,
    e.g.,~\cite{Zaafarani:2006:TITM}).}
    \label{fig:counter_rotations_ebsd}
\end{figure}

\begin{figure}[h!]
  \begin{center}
  \hspace{0.5cm}
  \subfigure{
    \includegraphics[width=8cm]{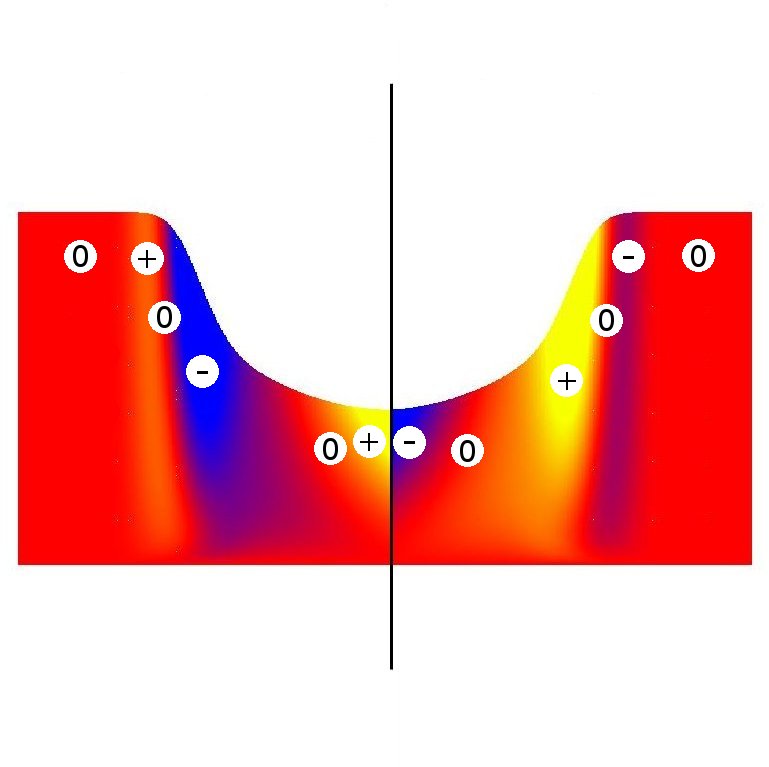}
    \raisebox{1.7cm}{
      \includegraphics[height=5cm]{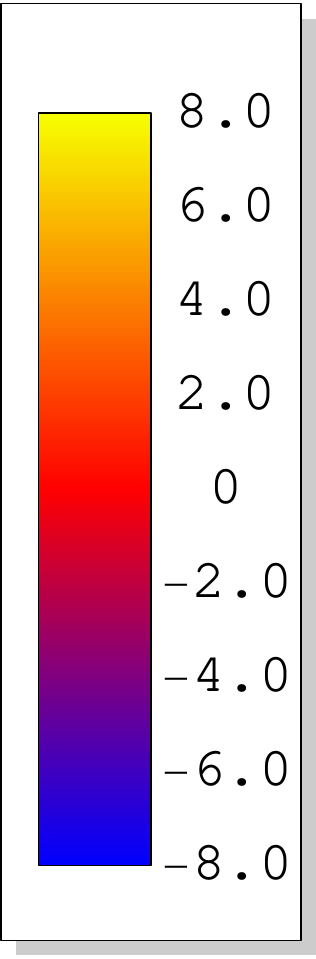}}
  }
  \end{center}
\vspace{-1cm}
\caption{Planar spins $\alpha_{e_2}(\rpolar^\pm_{1,0}(F))$ for the
  synthetic nanoindentation mapping $\varphi_{\rm nano}$ evaluated
  on the intersection with the $y = 1/2$-plane. Note that both
  branches are patched together at $x = 0$ in order to illustrate
  the possibility of counter rotations. In the full boundary value
  problem~\eqref{eq:variational_formulation}, such discontinuities
  can be resolved by solutions with a steep gradient in the
  Cosserat microrotation field.}
  \label{fig:planar_spin_patched}
\end{figure}

For comparison, we show the planar spin $\alpha:\SO(3) \to (-\pi,\pi]$
in the $xz$-plane which is oriented such that the sign
of the rotation angles coincides with the usual convention
for angles in the plane. We have computed this spin for the
classical continuum rotation $\polar(F_\text{nano})$ and the
relaxed polar factors $\rpolar_{1,0}^\pm(F_\text{nano})$
in \figref{fig:planarSpins}. Both plots show values attached
to the deformed configuration. In order to indicate the rotation
patterns that can be realized by a solution of a Cosserat
boundary value problem, we have patched together the two
branches of $\rpolar_{1,0}^\pm(F)$ in a suitable way,
see~\figref{fig:planar_spin_patched}. Note that the discontinuity
in $x = 0$ can be resolved by a steep gradient in the
microrotation field $R$ which is highly localized. Similarly,
we expect that more complicated rotation patterns can be
realized as is typically expected from a diffuse interface
model.\footnote{The Cosserat curvature contribution can be
  interpreted as a regularization term in a diffuse interface
  approach for the field of microrotations $R$.}

To ease the comparison with the experimentally obtained lattice
misorientation angles depicted in \figref{fig:raabe_indent_fig13},
we have used a similar color map with the same domain $[-8^\circ,8^\circ]$.

We summarize that the relaxed-polar mechanism $\rpolar_{1,0}^\pm(F)$
is capable to produce non-classical rotation patterns
as opposed to the classical macroscopic continuum rotation
$\polar(F)$. Whether this mechanism can be successfully exploited
for the realistic modelling of nanoindentations is open to future
investigation. It seems to us that one cannot
expect to obtain realistic results on the nanometer scale without
incorporating the geometry of the crystalline material explicitly.
However, the energy-optimal deformation modes related to
the plane of maximal stretch $\mathrm{P}^{\rm ms}(F)$,
see~\figref{fig:plane_of_maximal_strain} are very interesting
on the right mesoscale. We imagine that the movement in glide
planes is averaged at coarse enough length scales and that the
material deforms rather freely in an arbitrary direction of maximal
planar loading, although the deformation itself is realized by a
complex super-position of dislocation glides respecting the crystal
mechanics specific to the crystalline material.

\countres
\section{Conclusion}
\label{sec:conclusion}
In a series of nanoindentation experiments in copper single crystals
described and analyzed in~\cite{Zaafarani:2006:TITM,Zaafarani:2008:ODRP,Zaafarani:2008:ETIN,Demir:2009:IISE}, a formation of multiple typical
zones of lattice rotations was observed. These rotation patterns are
characterized by alternating signs of the field of lattice
misorientation angles. We are mostly considered with orientation
angles in the $\milPlane{1 1 \bar{2}}$ crystal plane below the
indentation profile. Physically based crystal plasticity simulations
based on a model proposed in~\cite{Ma:2004:CMFCC,Ma:2006:MCPM} reproduced
the rotation and pile-up patterns below the indentation profile quite
well~\cite{Zaafarani:2006:TITM,Zaafarani:2008:ETIN}.
The computed lattice misorientation angles were quantitatively
slightly overestimated, but possible strategies to overcome this
phenomenon by an improved treatment of dislocation effects are
mentioned.

For the sole purpose of demonstrating failure of the method,
Zaafarani carried out simulations based on an isotropic $J_2$-plasticity
model~\cite{Zaafarani:2008:ETIN}. As expected, this approach did not
reproduce the experimentally observed rotation zones. This motivated us
to investigate whether an \emph{isotropic} Cosserat model can produce
more realistic rotation patterns.

In the present work, we have made another step forward in our
ongoing analysis of the geometry of the isotropic geometrically
nonlinear Cosserat model~\eqref{eq:variational_formulation}
with zero Cosserat couple modulus $\mu_c = 0$ which was originally
proposed in the habilitation
thesis of the second author~\cite{Neff:2004:GECBT}.
Our analysis shows that the Cosserat model with
\mbox{$\mu_c = 0$} \emph{is capable to produce non-classical
rotation patterns} which are at least similar to observations
in nanoindentation experiments~\cite{Zaafarani:2006:TITM}.
This is only possible due to the additional field of Cosserat
microrotations $R: \Omega \to \SO(3)$. Note that the understanding
of the effects that can be generated by the field of microrotations $R$
is an important aspect of our efforts towards a
deeper understanding of the variational Cosserat model. For now,
we conclude that the energy-minimizing rotation
fields $\rpolar_{\mu,0}^\pm(F)$, i.e., with $\mu_c = 0$,
do provide an interesting new mechanism.

Due to the isotropic modelling approach, one cannot
expect to obtain realistic results representing the rich
anisotropic material behavior which a copper crystal lattice
can produce at the nanometer scale.\footnote{Note that it is possible
to modify the Cosserat model such that it includes anisotropic
effects due to a particular crystal lattice structure.} Our present
analysis of a synthetic nanonindenation and the direct comparison
with measurements in a two-dimensional section shows
that the rotation fields $\rpolar_{1,0}^\pm(F_{\rm nano})$
produce an interesting match, if we compose the two branches of
$\rpolar_{1,0}^\pm(F)$ in a suitable way. This composition is not
superficial, since in the solution of a Cosserat boundary
value problem, a mixture of the two branches is the \emph{expected}
result. In fact, such complex rotation patterns have already been
observed in numerical Cosserat plasticity simulations by
Blesgen, see~\cite{Blesgen:2012:DPLC,Blesgen:2013:DPF,Blesgen:2014:DPC,Blesgen:2015:RDZ}.

We briefly report on possible future directions. It remains to be
seen how the relaxed-polar mechanism discovered in our
investigation of the perfectly spatially decoupled model with
$\Lc = 0$ in~\cite{Neff_Biot07,Fischle:2016:OC2D,Fischle:2016:OC3D}
is reflected in solutions to Cosserat boundary value problems.
We also expect non-classical rotation patterns with essential
features of $\rpolar_{1,0}^\pm(F)$ in numerical solutions in both
two and three space dimensions.
It is an interesting question, whether these deviations are
also realized in numerical simulations of the Cosserat model
with $0 < \Lc \ll 1$ (or even observable in experiments such
as nanoindentation). This raises questions on how the solutions
to the Cosserat boundary value problem with $\Lc = 0$ and
$0 < \Lc \ll 1$ are precisely related. This is still an open
problem. We expect that the relaxed-polar mechanism which stems
from the investigation of the limit case without Cosserat curvature
$\Lc = 0$ plays an important part also for small length scales
$0 < \Lc \ll 1$.

Let us conclude this contribution with an interesting experimental
observation with possible implications on the choice of material
parameters in the Cosserat model.

\begin{rem}[Counter-rotations at the onset of nanoindenation and zero
    Cosserat couple  modulus $\mu_c = 0$]
  The counter-rotation patterns forming during nanoindentation
  experiments are quite sensitive: they form right at the onset
  of the process, i.e., for very small indenter loads.
  A zero Cosserat couple modulus $\mu_c = 0$ introduces this kind of
  sensitivity in the Cosserat model as opposed to a strictly positive
  $\mu_c > 0$. In the latter case, the non-classical effects are only
  triggered once a certain threshold for the maximal mean planar
  strain $\smp(F)$ is exceeded~\cite{Fischle:2016:OC2D,Fischle:2016:OC3D}.
\end{rem}

\section{Acknowledgments}
A. Fischle is currently supported by German Research Foundation (DFG)
grant SA2130/2-1 and has been partially supported by DFG grant
NE902/2-1 (also: SCHR570/6-1). We thank our anonymous
  referees for their concise and inspiring remarks which helped us
  to improve the paper.
 \countres
\FloatBarrier

\bibliographystyle{plain}

{\footnotesize
  \setlength{\bibsep}{1.25pt}
  \bibliography{./finera-2016-nano2d-arxiv-v2}
}

\appendix
\section{Supplementary Material}
\label{sec:appendix}
{\footnotesize
\subsection{Best-approximation in $\Psym(3)$ (due to Higham)}
First, we provide a short exposition of Higham's best approximation of
a given matrix $X \in \R^{3 \times 3}$ in the closed cone of symmetric
positive semi-definite matrices $\closure{\Psym(3)}$.
\begin{theo}[Higham's approximation theorem~\cite{Higham:1988:NSPSD}]
  Let $X \in \R^{3 \times 3}$ and let
  \begin{equation}
    \pi(X) \;\eqdef\; \arginf{S \,\in\, \closure{\Psym(3)}}{\hsnorm{X - S}}
  \end{equation}
  denote the mapping of $X$ to its best approximation in $\closure{\Psym(3)}$
  with respect to the Frobenius matrix norm. Then there exists a
  unique best approximation $\pi(X) = P \in \closure{\Psym(3)}$
  of the explicit form
  \begin{equation}
    \pi(X) \;=\; \frac{1}{2}\left(\sym(X) + Z\right)\;.
  \end{equation}
  Here $Z \in \closure{\Psym(3)}$ is the symmetric factor of the
  polar decomposition $\sym(X) = Q\,Z$, $Q \in \O(3)$.
\end{theo}

\begin{cor}[The symmetric positive definite case]
  If the symmetric part $\sym(X)$ is positive definite, then
  \begin{equation}
    \pi(X) \; = \; \sym(X)\;.
  \end{equation}
\end{cor}
\textit{Proof.}
We apply Higham's approximation theorem. First we compute the right polar
decomposition of $\sym(X) = Q\,Z$ which yields the symmetric factor
$Z \in \Sym(3)$. Similar to the usual computation of the stretch tensor, we
find
\begin{equation}
  Z = \sqrt{\sym(X)^T\sym(X)} = \sqrt{\sym(X)^2}\;.
\end{equation}
By assumption $\sym(X)$ is symmetric positive definite. Thus, since the symmetric positive
definite matrix square root is unique, we obtain
\begin{equation}
  Z = \sqrt{\sym(X)^2} = \sym(X)\;.
\end{equation}
This shows that the best approximation of $X$ is given by
\begin{align}
  \pi(X) = \frac{1}{2}\left(\sym(X) + \sym(X) \right) = \sym(X) \in \Psym(3)\;.\tag*{$\blacksquare$}
\end{align}

Next, we present a simple condition which implies the positive definiteness
of $\sym(X)$. Note first that
\begin{equation}
  \scalprod{\sym(X)\xi}{\xi} = \scalprod{X\xi}{\xi} > 0
  \;\isequivto\;
  \scalprod{(X - \id +\id)\xi}{\xi}
  = \underbrace{\scalprod{\sym(X - \id)\xi}{\xi}
    + \norm{\xi}^2}_{\rm A} > 0\;.
\end{equation}

Since
\begin{equation}
  \scalprod{(X - \id)\xi}{\xi} = \scalprod{\sym(X - \id)\xi}{\xi} \leq \hsnorm{\sym(X - \id)}\norm{\xi}^2\;,
\end{equation}
we observe that
\begin{equation}
  \scalprod{\sym(X)\xi}{\xi} = {\rm A} \geq \norm{\xi}^2 - \hsnorm{\sym(X - \id)}\norm{\xi}^2
= \left(1 - \hsnorm{\sym(X - \id)}\right)\norm{\xi}^2\;.
\end{equation}

We conclude that $\sym(X)$ is positive definite if
$\hsnorm{\sym(X - \id)} < 1$. Moreover, the inequality
$\hsnorm{\sym(X - \id)} \leq \hsnorm{X - \id}$
shows that $\sym(X)$ is positive definite if
$\hsnorm{X - \id} < 1$.

\subsection{Algorithmic implementation of $\rpolar_{1,0}^{\pm}(F)$}
For our computations, we have used the following implementation
of $\rpolar_{1,0}^{\pm_Q}(F)$ in \mathematica:

\begin{verbatim}
(* Right polar decomposition of the deformation gradient F*)
PolarDecomposition[F_] := Module[{PolarF, U},
   U = MatrixPower[Transpose[F].F, 1/2];
   PolarF = F.Inverse[U];
   {PolarF, U};
];

(* Relaxed polar factors $\rpolar_{1,0}^\pm(F)$ of the deformation gradient F *)
RelaxedPolarDecomposition[F_] :=
  Module[{Q, Spec, RHat, alphaHat, PolarF, RPolarPlus, RPolarMinus},

   {PolarF, U} = PolarDecomposition[F]},
   (* The columns of Q are the eigenvectors of U in the canonical basis *)
   Q = Transpose[Map[#/Sqrt[#.#] &, Eigenvectors[U]]];

   (* Enforce (nearly) consistent orientation of orthonormal eigenframe:
      this is a specific solution for the presented synthetic nanoindentation. *)
   If[Det[Q] < 0,
    Q[[All, 3]] = -Q[[All, 3]];
    ];

   (* Singular value spectrum of F / Eigenvalues of U ordered by decreasing magnitude *)
   Spec = Eigenvalues[U];

   (* Maximal mean planar strain *)
   S_MMP = 1/2*(Spec[[1]] + Spec[[2]]);

   (* Compute relative rotation angle for $\mu = 1$ and $\mu_c = 0$ *)
   If[S_MMP > 1, alphaHat = ArcCos[1/S_MMP], alphaHat = 0];

   (* Parametrize relative rotation $\hat{R}$ in plane of maximal stretch *)
   RHat[alphaHat_] := {{Cos[alphaHat], -Sin[alphaHat],  0},
                       {Sin[alphaHat],  Cos[alphaHat],  0},
                       {            0,              0,  1}};

   (* Positive branch relative to a choice of orientation for Q *)
   RPolarPlus = PolarF.Q.RHat[-alphaHat].Transpose[Q];

   (* Negative branch relative to a choice of orientation for Q *)
   RPolarMinus = PolarF.Q.RHat[alphaHat].Transpose[Q];

   (* Return both branches *)
   {RPolarPlus,RPolarMinus}
];
\end{verbatim}
}
 \countres

\end{document}